\documentclass[sn-mathphys]{sn-jnl}% Math and Physical Sciences Reference Style
\jyear{2021}%

\theoremstyle{thmstyleone}%
\theoremstyle{thmstyletwo}%

\theoremstyle{thmstylethree}%

\usepackage{amsfonts,amssymb}
\begin{document}

\title[Discrete lattices with non-local interactions and non-convex energy]{Thermal control of nucleation and propagation transition stresses in discrete lattices with non-local interactions and non-convex energy}


\author[1,2]{\fnm{Andrea} \sur{Cannizzo}}\email{andrea.cannizzo@iemn.fr, https://orcid.org/0000-0002-7616-156X}

\author[2]{\fnm{Luca} \sur{Bellino}}\email{luca.bellino@poliba.it,  https://orcid.org/0000-0002-4823-3589 }

\author[3,4]{\fnm{Giuseppe} \sur{Florio}}\email{giuseppe.florio@poliba.it, https://orcid.org/0000-0002-5499-2530}

\author[3]{\fnm{Giuseppe} \sur{Puglisi}}\email{giuseppe.puglisi@poliba.it, https://orcid.org/0000-0002-6771-5495}

\author*[1]{\fnm{Stefano} \sur{Giordano}}\email{stefano.giordano@univ-lille.fr, https://orcid.org/0000-0003-4023-5384}

\affil*[1]{Univ. Lille, CNRS, Centrale Lille, Univ. Polytechnique Hauts-de-France, UMR 8520 - IEMN - Institut d'Electronique de Microélectronique et de Nanotechnologie, F-59000 Lille, France}

\affil[2]{Politecnico di Bari, (DMMM) Dipartimento di Meccanica, Matematica e Management, Via Re David 200, I-70125 Bari, Italy}

\affil[3]{Politecnico di Bari, Dipartimento di Ing. Civile, Ambientale, del Territorio, Edile e di Chimica, Via Re David 200, 70126 Bari, Italy}

\affil[4]{INFN, Sezione di Bari, I-70126, Italy}

\abstract{Non-local and non-convex energies represent fundamental interacting effects regulating the complex behavior of many systems in biophysics and materials science.
We study one dimensional, prototypical schemes able to represent the behavior of several biomacromolecules and  the phase transformation phenomena in solid mechanics.
To elucidate the effects of thermal fluctuations on the non-convex non-local behavior of such systems, we consider three models of different complexity relying on thermodynamics and statistical mechanics: (i) an Ising-type scheme with an arbitrary temperature dependent number of interfaces between different domains, (ii) a zipper model with a single interface between two evolving domains, and (iii) an approximation based on the stationary phase method.
In all three cases, we study the system under both isometric condition (prescribed extension, matching with the Helmholtz ensemble of the statistical mechanics) and isotensional condition (applied force, matching with the Gibbs ensemble).
Interestingly, in the Helmholtz ensemble the analysis shows the possibility of interpreting the experimentally observed thermal effects with the theoretical force-extension relation characterized by a temperature dependent force plateau (Maxwell stress) and a force peak (nucleation stress).
We obtain explicit relations for the configurational properties of the system as well (expected values of the phase fractions and number of interfaces).
Moreover, we are able to prove the equivalence of the two thermodynamic ensembles in the thermodynamic limit.
We finally discuss the comparison with data from the literature showing the efficiency of the proposed model in describing known experimental effects.}

\keywords{configurational transitions, phase transformations, statistical mechanics of bistable systems, nanowires}



\maketitle

\section{Introduction}

Several natural and artificial systems, as typically encountered in biology and modern nanotechnology, exhibit a combination of fundamental effects of non-locality and non-convexity, resulting in a variety of rather complex physical responses \cite{prote,prote1,res,res1}.
A deep understanding of these phenomena is therefore essential for the analysis and design of such systems.
The non-convexity feature is related to the possibility that the potential energy of the system may have different basins leading to wiggly energy landscapes and possibly many competing metastable states.
Often, these systems are composed of several units (which may be identical, or inhomogeneous) and each unit is characterized by a bistability (or in general a multistability).
This assumption describes the possibility for each unit to be in two (or more) distinct states.
In this framework, non-locality describes the possibility that the units can be in strong interaction with each other and thus the state in which one unit is found affects the state of the others, particularly those that are spatially closer.
In the biological context this form of interaction is sometimes referred to as cooperativity \cite{coop} and may be fundamental in many crucial biological and medical phenomena such as protein folding/unfolding, DNA degradation and resulting diseases \cite{GGBP,pinzette}.
The complex response of such systems is the result of the energy competition among the many metastable configurations, regulated by the 
temperature controlled exploration of the overall energy landscape. We show how three different statistical mechanics approaches, taking care of the elastic properties of the different configurations, can give important insights for many observed physical and biophysical phenomena.

Many examples can be theoretically inscribed into the previously introduced conceptual framework.
For example, in the biological field we may consider the conformational transitions in bio-polymeric chains \cite{mol1,mol2,mol3,rief3,mol4,mol5,mol6,mol7,mol8,mol9,mol10}, and the sarcomeres behavior in skeletal muscles \cite{HS,hill,epstein,caruel1,caruel2,caruel-ropp,caruelast,caruelastlast}.
On the other hand, concerning artificial systems, we can think to waves propagation in bistable lattices \cite{rafsa,daraio,bgtb,wbl,sefi,berto}, energy harvesting through multistable chains \cite{hwa,harne1,harne2}, and the plasticity and hysteresis in phase transitions and martensitic transformations of solids \cite{eri,ball,zan,mullerv,shaw,abe,bengiv,tri,col,ren,muller,fed,pt1,pt2,tvai,p1,p2}.

In this paper, we are interested in the observed thermal effects in previous examples, that become increasingly important as the size of the system decreases. The analysis is therefore relevant in systems of nanoscopic dimensions \cite{dmo0c} or when, as in biological and polymeric soft matter, the competing contributions are of entropic type with small energy differences and low barriers \cite{aes}.
For this reason, in order to have a correct physical description of the static and dynamic features of these systems, it is not sufficient to rely on classical discrete or continuous mechanics, but we must take into account the equilibrium or non-equilibrium statistical mechanics.
So doing, we can obtain important information on the thermodynamic picture of the folding/unfolding process in a macromolecular chain or a detailed description of the thermal effects on the microstructure evolution in a two-phase solid material.

Concerning materials science, and in particular the mechanics of solids for multiphase materials, many approaches can be found in the literature to describe the microstructure evolution.
The Ericksen pioneering work proposed a variational energetic approach in the context of non-linear continuum elasticity theory with non-convex energy densities \cite{eri}.
This methodology has been further generalized to describe phase transformations at the microstructure level \cite{ball,zan,mullerv}.
Nevertheless, these continuum variational approaches neglect interfacial energy effects and non-local interactions, which are crucial contributions for the description of the realistic microstructure evolutions \cite{shaw}.
As a matter of fact, the minimization
of the non-convex elastic energy without non-local interactions cannot completely describe the  nucleation and propagation of finite domains \cite{abe,bengiv}.
Therefore, surface energy contributions have been introduced by means of higher gradient energy terms \cite{tri,col}, and by non local interactions \cite{ren}.
A similar research line, in the context of the discrete mechanics, has been developed from the pioneering work of M\"{u}ller and Villaggio \cite{muller}.
The basic model is composed of a one-dimensional lattice of units with a non-convex potential energy and an intrinsic length-scale \cite{fed}.
This scheme allows the description of energy barriers, metastable states, quasi-plastic and pseudo-elastic behaviors \cite{pt1,pt2}.
Also in this context, the model has been extended with non local energy terms able to capture the different features of phase nucleation and propagation \cite{tvai}.
Further generalizations consider the influence of boundary conditions, enabling a more detailed identification of the internal and boundary phase nucleations \cite{p1,p2}.
Recent models investigate the austenite to martensite phase transitions in wires, eventually describing the shape memory alloy behavior under uniaxial tension \cite{duval,alessi,zilong}.

Many similar methods have been elaborated to describe the configurational transitions in biological macromolecules (mainly proteins), undergoing folding/unfolding processes.
In particular, those theories are able to explain the saw-tooth-like force-extension response observed in several experiments.
A model has been proposed for macromolecules unfolded in atomic force microscopes, and validated for titin and RNA hairpins \cite{staple}.
An approach based on the equilibrium statistical mechanics is based on a Landau-like free energy and predicts a sequence of first-order phase transitions in correspondence to the unfolding processes \cite{prados1,prados2}.
Further investigations are based on the energy minimization of a bistable system and agree with the pattern observed in titin experiments \cite{saccomandi}.
Also the Monte-Carlo implementation of a two-state theory for single-molecule stretching experiments has been proposed \cite{manca3}.
Finally, the mechanical unfolding of proteins has been also modeled through domains interactions described by the Ising model \cite{maka,PREnew}.

In order to properly introduce the thermodynamics on non-convex, non-local systems, we adopt the method of the spin variables, by extending classical one-dimensional schemes to consider the fundamental effect of stiffness and elasticity of the different states.
The first theoretical ideas introducing this technique can be found in the early models of the bio-mechanical response of skeletal muscles \cite{HS,hill}.
This technique has been further generalized to study different multistable systems \cite{caruel1,caruel2,caruel-ropp,caruelast,prr,prr_,aes}, and macromolecular chains \cite{soft,JCPnew,physa,giobene,prot,prot1}.
This approach is based on the introduction of a series of discrete variables (the so-called spin variables), which are able to identify the state of the units.
In other words we associate to each unit a sort of ``bit'', identifying the folded or unfolded state of the unit.
So doing, we can consider two separated and different quadratic functions representing the wells of the potential energy, instead of the more complicated original bistable function.
The introduction of the spin variables strongly simplifies the calculation of the partition functions and, consequently, the analysis of the macroscopic thermodynamic quantities.
Indeed, in order to calculate these partition functions, we sum over the spin variables and we integrate the classical continuous variables.
Since the separated wells are represented by quadratic terms,   the integration can be performed straightforwardly since it acts on Gaussian functions.
This theoretical approach is therefore able to yield closed form results useful to better understand the underlying physics.
Moreover, by means of this technique we can study different ensembles of the statistical mechanics, corresponding to different mechanical boundary conditions.
From one hand, we can analyze the behavior of the system under an applied force (isotensional condition), corresponding to the Gibbs ensemble.
On the other hand, we can also investigate the features of the system with prescribed extension (isometric condition), corresponding to the Helmholtz ensemble.
As described in Ref.\cite{prot}, these boundary conditions can be considered as limiting configurations of the more realistic case of an elastic device interacting with the system.
Moreover, from a theoretical point of view, 
the comparison of the force-extension response within different ensembles is an important point useful to understand the concept of ensembles equivalence in the thermodynamic limit, as discussed below.
The spin variable method has been successfully used to study the nanomechanics of macromolecules \cite{soft,JCPnew,physa,giobene,prot,prot1}, the denaturation of macromolecules \cite{dena1,dena2}, the non-local or cooperative effects \cite{PREnew,aes}, skeletal muscles \cite{caruel1,caruel2,caruel-ropp,caruelast}, and systems with transitions between unbroken and broken states \cite{robin,truskinovskypar1,truskinovskypar2,prr,prr_}.

\begin{figure}[t!]
\centering
{\includegraphics[scale=0.8]{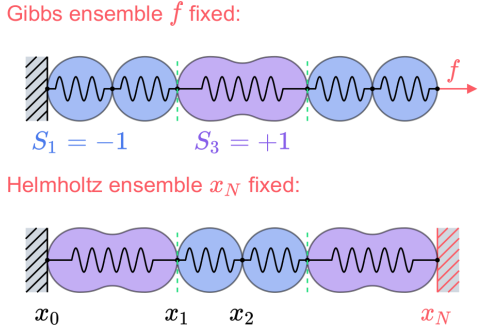}}
\caption{\label{fig:BSPE} Scheme of the discrete system with variable phase configuration.  Top panel: isotensional loading (Gibbs ensemble) of the chain with applied force $f$. Bottom panel: isometric loading (Helmholtz ensemble) of the  chain with the prescribed extension $x_N$. Each chain unit, in both ensembles, feels an Ising interaction energy $J$ due to the nearest neighbors. As an example, in both Gibbs and Helmholtz schemes, we show two interfaces between folded and unfolded domains.}
\end{figure}

\begin{figure}[t!]
\centering
{\includegraphics[scale=0.9]{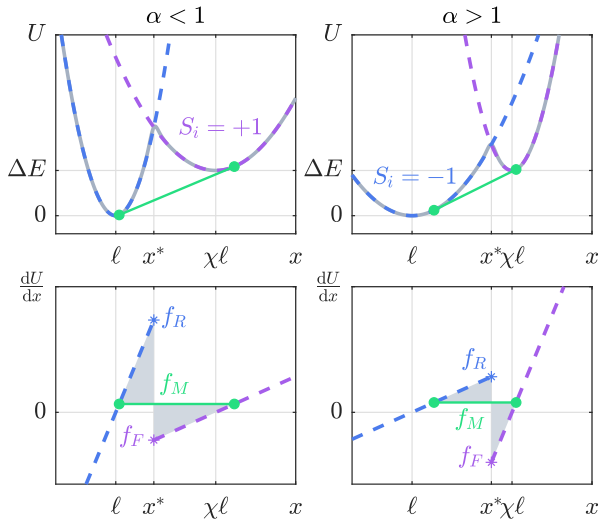}}
\caption{\label{fig:wells} Elastic behavior of the chain units. Top panels: bistable potential energy $U=U(x)$ versus $x$ for each unit, where $x=x_i-x_{i-1}$. The folded well corresponds to $S_i = -1$ and it is characterized by $L(-1)=\ell$, $Q(-1)=0$, $K(-1)=k$. The unfolded well, identified by $S_i = +1$, is defined by $L(+1)=\chi\ell$, $Q(+1)=\Delta E$, $K(+1)=\alpha k$.  Bottom panels: we show the quantity $\mathrm{d}U/\mathrm{d}x$ versus $x$, exhibiting the characteristic force jump. Both the cases with $\alpha<1$ and $\alpha>1$ are shown on the left and on the right, respectively. In all plots, the solid green lines represent the transition paths (at $T=0$), discussed in the main text.}
\end{figure}

Here, we analyze a discrete chain with $N$ bistable units (see Fig.\ref{fig:BSPE}), where the two potential energy wells of each unit are characterized by different elastic constants $k$ (pristine or folded state) and $\alpha k$ (extended or unfolded state), where $\alpha > 0$ (see Fig.\ref{fig:wells}).
In addition, the two states are separated by an energy jump $\Delta E$, representing the (Helmholtz) transition energy, and have their equilibrium lengths equal to $\ell$ and $\chi\ell$, with a length rise of $\Delta x=\ell(\chi-1)$, where $\chi>1$ (see Fig.\ref{fig:wells}).
The assumption of considering two different elastic constants of the wells, already considered in the purely mechanical case in Ref.\cite{pt1}, involves important novelties compared to the case with identical constants.
Indeed, when the folded and unfolded elastic constants are equal ($\alpha=1$), the conformational transitions correspond to a temperature independent average plateau force \cite{aes,soft,JCPnew,physa,giobene,prot,prot1}.
This result can be simply explained in the framework of the Bell relation $f=\Delta E/\Delta x$, discovered in the context of  cell adhesion \cite{bell1,bell2,manca3}.
This plateau force, which depends neither on the spring constant nor on the temperature $T$, can be explained as follows.
We consider two potential energies $U_f(x)=\frac{1}{2}k(x-\ell)^2-fx$ and $U_u(x)=\Delta E+\frac{1}{2}k(x-\chi\ell)^2-fx$, corresponding to the folded and unfolded states of the unit under force $f$ when $\alpha=1$.
In both cases, the equilibrium lengths are defined by $\partial U_f/\partial x=0$ and $\partial U_u/\partial x=0$ and we get $x_f=\ell+f/k$ and $x_u=\chi\ell+f/k$.
Finally, the unfolded configuration is more favorable than the folded one when $U_u(x_u)<U_f(x_f)$, which corresponds to $f>{\Delta E}/{\Delta x}$. The quantity $f_M={\Delta E}/{\Delta x}$ is the so-called Maxwell force for the case with $\alpha=1$.
This approach can be easily generalized to the case with $\alpha\neq 1$ and $T=0$ (purely mechanical behavior). The same analysis yields in fact the following quadratic equation for the transition force $f_M$
\begin{equation}
\label{zerotemp}
    f_M^2(1-\alpha)+2\alpha k f_M \Delta x-2\alpha k \Delta E=0,
\end{equation}
where $\Delta x=\ell(\chi-1)$, as before. This value of force has two important properties (both valid for $\alpha<1$ and $\alpha>1$): (i) on the plane $(U,x)$ it is represented by an inclined straight line that is the common tangent to the two parabolas of the wells (see solid green lines in Fig.\ref{fig:wells}, top panels); (ii) on the plane $(\mathrm{d}U/\mathrm{d}x,x)$ is represented by a horizontal straight line (see solid green lines in Fig.\ref{fig:wells}, bottom panels) that makes equal the two areas of the indicated triangles (see shaded regions in Fig.\ref{fig:wells}, bottom panels).  This last property gives the name Maxwell force to the transition force because the equality of the two triangles is reminiscent of Maxwell's construction on the isothermal Van der Waals curves in the pressure-volume plane of a real gas \cite{prigo}. 
The situation becomes much more complicated when $\alpha\neq 1$ and $T>0$ and the transition force $f_M$, based on statistical mechanics analysis, is temperature dependent. Indeed, the asymmetry of energy wells makes entropic contributions crucial.
It means that a temperature dependent term must be added to Eq.(\ref{zerotemp}) when $\alpha\neq 1$ and $T>0$. 
This point is the focus of this paper and it is extensively examined in the following development.

Other two special values of the force are of interest in this papers and represent the maximum (roof) $f_R$ and the minimum (floor) $f_F$ values of the force, such that for $f\in(f_F,f_R)$ both the first phase and the second phase can exist. They can be simply obtained as
\begin{eqnarray}
  f_R&=&k \left(x^*-\ell\right), \label{fR}\\
  f_F&=&\alpha k \left(x^*-\chi\ell\right), \label{fF}
\end{eqnarray}
where $x^*$ represents the point of intersection of the two parabolas (where the top of the energy barrier is reached, see Fig.\ref{fig:wells}).

We also introduce in the discrete lattice of non-convex  elements (see Fig.\ref{fig:BSPE}) a non-local interaction described through an Ising scheme.
This feature is crucial to consider a form of cooperativity in the biological context or, equivalently, an interface energy between folded and unfolded domains in the materials science context.
The Ising interaction energy $J$ can be considered positive (cooperative case) when adjacent units prefer to be in the same state, and negative (anti-cooperative case) when they prefer to be in two different states. 
To focus mainly on the interesting temperature dependent force plateau behavior, in this study we will discuss only the ferromagnetic-like interactions, with $J>0$.
The important theoretical novelty with respect to classical spin models in physics is the fundamental role of elasticity and, in particular, the effect generated by the different elastic behavior of the two phases.
Furthermore, for these systems, we can determine the analytic expression of the partition function in both Gibbs and Helmholtz ensembles (isotensional and isometric conditions, as shown in Fig.\ref{fig:BSPE}). 
This can be done by means of an \textit{ad hoc} implementation of the transfer matrix technique for the Gibbs case \cite{baxter}, and by using the Laplace transform relationship between the partition functions of conjugated ensembles for the Helmholtz case \cite{weiner}. 

The first analysis, within the Gibbs ensemble, provides evidence that the force plateau  is temperature dependent and we obtain its expression in the limit of large values of $N$ and large (positive) values of the ratio $J/K_B T$ ($K_B$ being the Boltzmann constant), which represents the strongly ferromagnetic case in the thermodynamic limit.
Defining $\beta=1/K_B T$, we can introduce the quantity $\tilde{\beta}=J\beta$ that represents the competition between the Ising interaction energy and the entropic contributions.
Moreover, for isometric loading, in the Helmholtz ensemble, we find a peak force at the beginning of the plateau, representing the nucleation of a new domain with unfolded units. 
This is an important feature, typically observed in experimental measurements \cite{exp1,exp2,exp3,exp4,exp5,exp6}, and in molecular dynamics simulations of the microstructure evolution in nano-systems \cite{dmo0,dmo0b,dmo0c,dmo1,dmo2,dmo3,dmo3b,dmo4,dmo5,dmo6,dmo7,dmo8,dmo9,dmo10}. 
In this Ising model, the microstructure evolution of folded and unfolded domains is free, and regulated by the competition between interface energies and entropic contributions and, as we will show, the stress peak is an effect induced by the introduction of interface energy terms.
The number of interfaces may vary in the whole range between 0 and $N-1$ and typically it increases with the temperature \cite{aes}. 
The knowledge of the partition function allows a full analysis of the configurational properties of the system leading to the quantitative evaluation of the average number of unfolded units and the average number of interfaces. 
In particular, this allows us to observe that the microstructure evolution, under increasing extension, is characterized by a single moving domain wall between folded and unfolded regions only when $\tilde{\beta} \gg 1$, \textit{i.e.} when the system is strongly ferromagnetic, whereas for $\tilde{\beta} \ll 1$, i.e. when temperature increases, entropic energy terms favor solutions with an increasing number of interfaces.

An interesting point concerning these systems is the equivalence of the ensembles in the thermodynamic limit (\textit{i.e.}, for $N\to\infty$) \cite{winkler,manca1,manca2,manca4,cmat}. 
Two conjugated ensembles are said to be equivalent  when the macroscopic behavior described by the force-extension relation is the same for $N\to\infty$. 
In general, it is difficult to prove for a given system if two statistical ensembles are equivalent. 
Although there are some particular rules, there are no general criteria or theorems for determining whether a system satisfies such an equivalence \cite{manca4}.
Several examples of non-equivalence are well known in the literature \cite{ine1,ine2,ine3,ine4,ine5,ine6}. 
In our case, the analysis of the equivalence is rather difficult, mainly due to the overly complicated mathematical form of the Helmholtz partition function. 
But, since the systems with positive and sufficiently intense Ising interaction ($\tilde{\beta}\gg 1$) are the most interesting for practical applications, we can limit our analysis to the case of a system where the number of interfaces can take only the values 0 and 1. 
This observation is at the origin of the second approach proposed in the paper, called the zipper model, previously adopted in other statistical mechanics investigations \cite{zip1,zip2,zip3,zip4}. 
The main assumption is the analysis of solutions with none or one interface, which correspond to the previous Ising scheme only when $\tilde{\beta} \gg 1$. 
This simplification makes the thermodynamic limit analysis more transparent and, in this zipper case, we can explicitly prove the equivalence of the isotensional and isometric ensembles for $N\to\infty$. 
In addition, the result obtained under the zipper assumption, within the Helmholtz ensemble, can be further simplified by means of the stationary phase method (large values of $N$), leading to the third approach here discussed to describe non-convex discrete systems with non-local interactions. 
This final approximation is particularly useful since yields a compact mathematical expression for the force-extension curve and it allows the explicit calculation of the force peak in the Helmholtz ensemble, which is a crucial quantity in several experiments and numerical simulations, as discussed above. 
From one side, it leads to draw some comparisons between our theoretical results and  data from the literature; from the other side, it suggests both the possibility of designing new materials with required transition properties, and the possibility of controlling them through external thermal fields.

The structure of the paper is the following.
In Section II we introduce the Ising scheme and we discuss in detail the proposed solutions for the Gibbs and Helmholtz ensembles. 
In Section III we explain how to obtain the thermodynamic limit within the Gibbs ensemble and we discuss the important effect of temperature dependent force plateaux. 
In Section IV we introduce the zipper model: we discuss the Gibbs and Helmholtz ensembles and we draw a comparison with the previous Ising scheme. 
In Section V we study the approximation based on the stationary phase and we perform a detailed analysis of the first peak force within the Helmholtz ensemble. 
Finally, in Section VI, to support the obtained analytical results, we discuss some explicit, quantitative comparisons with data from the literature concerning the behavior of nanowires with microstructural evolution.

%-----------------------------------------------------------------------------------------------------
\section{Two-state chain with Ising non-local interactions}
\label{Isingmodel}

Consider a discrete chain of $N$ two-state elements, each described by a bistable potential energy (see Figs.\ref{fig:BSPE} and \ref{fig:wells}), that interact also non-locally.
We distinguish the two phases using the spin variable $S_i$ assuming values in $\{-1,+1\}$. In particular $S_i=-1$ corresponds to the first well (folded element), whereas $S_i=+1$ corresponds to the second well (unfolded element). While in a previous work the authors considered identical wells \cite{aes}, the main hypothesis of this work is that the two phases are characterized by two different elastic constants $K(S_i)$, associated with two natural lengths $L(S_i)$ and two basal energies $Q(S_i)$.  Specifically, we assume, without loss of generality, that $L(-1)=\ell$, $L(+1)=\chi\ell$, $Q(-1)=0$, $Q(+1)=\Delta E$, $K(-1)=k$, $K(+1)=\alpha k$, where $\alpha>0$, $\chi>1$ and $\Delta E$ is the energy jump between the states. 
In these hypotheses, and introducing Ising non-local interaction terms, the overall Hamiltonian assumes the compact form
\begin{equation}
\label{overallH}
H = \sum_{i=1}^{N}\left\lbrace Q(S_i)+\frac{K(S_i)}{2}\left[\left(\lambda_i-\lambda_0(S_{i})\right)\ell\right]^2\right\rbrace-J\sum_{i=1}^{N-1}S_{i}S_{i+1}.
\end{equation}
Here, the non dimensional parameter $\lambda_i=(x_i-x_{i-1})/\ell$ is the ratio between the $i$-th spring length and the folded rest length $L(-1)=\ell$, {i.e.} the spring stretch and $\lambda_0(S_{i})=L(S_i)/\ell$, {i.e.} the natural (zero-force) spring stretch. The parameter $J$ measures the non-local interaction strength. We remark that $J > 0$ corresponds to the ferromagnetic case, favoring phases coalescence.
It is useful to take into account the adimensional Hamiltonian $\tilde{H}$, obtained dividing $H$ by the interface energy $J$
\begin{equation}
\label{overallHa}
\tilde{H} = \sum_{i=1}^{N}\left\lbrace \tilde{Q}(S_i)+\frac{\tilde{K}(S_i)}{2}\left(\lambda_i-\lambda_0(S_{i})\right)^2\right\rbrace-\varpi\sum_{i=1}^{N-1}S_{i}S_{i+1},
\end{equation}
where $\tilde{Q}(S_i)=Q(S_i)/J$, $\tilde{K}(S_i)=K(S_i)\ell^2/J$ and $\varpi=1$ is a constant that we inserted in order to be able, later on, to calculate the average number of interfaces. We also introduce $\tilde k=k\ell^2/J$ so that $\tilde K(-1)=\tilde k$ and $\tilde K(+1)=\alpha \tilde k$.
We remark that the spin variables approach can be adopted only when we work not far from the thermodynamic equilibrium \cite{soft,giobene}. Indeed when rate effects are considered, the relaxation times of the system strongly depend on the energy barriers between the potential wells, which are neglected within our approach (see Ref.\cite{kramers} and recent generalizations in Refs.\cite{givli,givli1}). On the other hand, in the rate independent regime considered here, this approach allows us to describe non-convexity through the discrete parameters $S_i$, in the sense that a for fixed phases configuration ($S_i$, $i$=1,...,$N$), the energy is convex with respect to other (stretch) variables. This energy structure also ensures that the equilibrium solutions represent local elastic energy minima (metastable equilibrium states). It is also important to remark that the application of the spin variable approach is correct only when the energy barrier between the two states is sufficiently larger than the thermal energy $K_B T$.

For completeness, in Appendix \ref{appA}, we prove that the Hamiltonian function given in Eq.\eqref{overallH}, based on Ising interactions among spins, can be obtained through an approximation of the Hamiltonian function defined by non-local next-nearest-neighbor (NNN) elastic interactions. Observe that this is coherent with the results in Refs.\cite{p1,p2}, where the author shows that for small values of $J$ the non-local interaction energy is, as in the case of the Ising model, proportional to the number of interfaces plus possible higher order boundary energy terms (that are neglected in this paper). This is an important point since some previous investigations on thermal effects for multi-stable lattices considered NNN interactions \cite{aes}. The resulting analysis was complex enough to prevent a fully analytical solution, whereas the scheme with Ising interactions  proposed here represents an important step forward because the results are obtained in closed form.   

Here we analyze separately the two cases of assigned force (Gibbs ensemble) and assigned displacement (Helmholtz ensemble). We remark that detailed computations are reported in the Appendix \ref{appB}.

%----------------------------------------------------------------------------
\subsection{Ising model within the Gibbs ensemble}
\label{Gibbs}

The statistical mechanics in the case of assigned force $f$, within the Gibbs ensemble  (see Fig.\ref{fig:BSPE}, top panel), can be introduced by calculating the canonical partition function
\begin{equation}
Z_G(f)=\sum_{\{S_i\}}\int_{\mathbb{R}^{N}} e^{-\beta  \left(H- fx_N\right)} \mathrm{d}x_1 \dots\mathrm{d}x_N.
\label{dimension_Z_G}
\end{equation}
Due to the fact that it is more useful to consider dimensionless parameters, in Eq.\eqref{dimension_Z_G} we substitute $x_i$ with the stretch $\lambda_i$ and introduce the dimensionless force $\tilde{f}=f\ell/J$ and energy $\tilde{H}=H/J$ together with the main non dimensional parameter of the paper $\tilde{\beta}=\beta J$. We end up with the following partition function expression
\begin{equation}
Z_G(\tilde{f})=\ell^N\sum_{\{S_i\}}\int_{\mathbb{R}^{N}} e^{-\tilde{\beta} \left(\tilde{H}- \tilde{f}\,\sum_{i=1}^N\lambda_i\right)} \mathrm{d}\lambda_1 \dots\mathrm{d}\lambda_N.
\label{gpfa}
\end{equation}
The sums over $\{S_i\}$  are to be considered extended to the values $+1$ and $-1$ for each spin variable ($i=1,\dots,N$). Moreover, we have that $\sum_{i=1}^N\lambda_i\ell=x_N$ (where, without loss of generality, to avoid rigid motions, we have assumed $x_0=0$).
Following the calculations in Appendix~\ref{appB}, we obtain
\begin{equation}
Z_G(\tilde{f})	=\frac{\ell^N}{2\cosh\tilde{\beta}}\left[{\hat \lambda_1}^N +{\hat \lambda_2}^N+e^{-2\tilde{\beta}}\left({\hat \lambda_1}^N -{\hat \lambda_2}^N\right)\frac{{\hat \lambda_1}+{\hat \lambda_2}}{{\hat \lambda_1}-{\hat \lambda_2}}\right],
\label{ZGwithlambdas}
\end{equation}
where $\hat \lambda_{1,2}$ (${\hat \lambda_1}>{\hat \lambda_2}$) are the eigenvalues of the transfer matrix defined in Eq.~(\ref{eigvalT}) (see Appendix~\ref{appB}).
This result is similar to the one obtained in Ref.~\cite{PREnew}, where however the elastic constants were considered equal ($\alpha=1$), and where a three-dimensional structure was studied to deal with polymeric cooperative systems. 

\begin{figure}[t]
    \centering
    \includegraphics[scale=1]{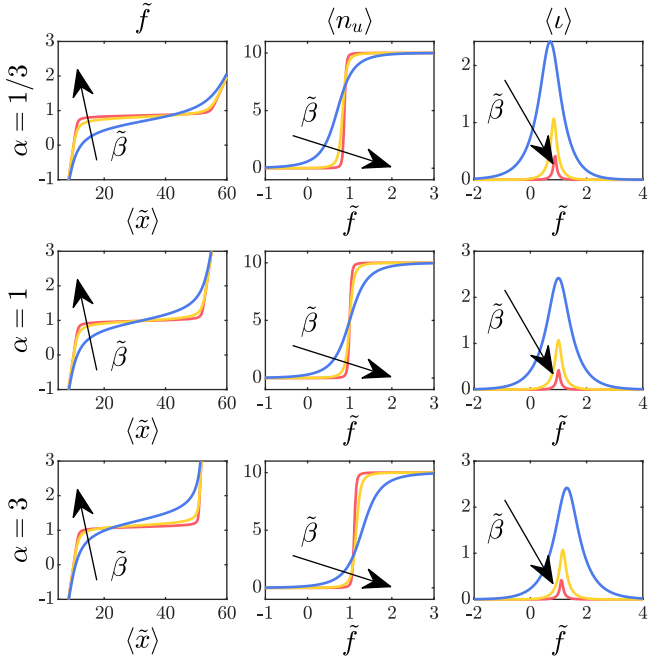}
    \caption{Behavior within the Gibbs ensemble with variable Ising coefficient $\tilde{\beta}$ and parameter $\alpha$. The quantities $\langle \tilde{x}\rangle$, $\langle n_u\rangle$ and $\langle \iota\rangle$ are represented versus the dimensionless force $\tilde{f}$, with a variable Ising coefficient $\tilde{\beta}=\{1/2, 1, 3/2\}$. In the first row, we used $\alpha=1/3$, in the second one $\alpha=1$, and in the third one $\alpha=3$. We adopted the parameters $N=10$, $\tilde{k}=6$, $Q(-1)=0$, $Q(+1)=\Delta \tilde{E}=4$, $\chi=5$.}
    \label{fig:ising1DdiffwellG}
\end{figure}

The knowledge of the Gibbs partition function allows us to calculate the expected value $\langle  x_N\rangle$ of the chain  length (\textit{i.e.}, the average value of the last position $x_N$, that from now on we rename $x$ for simplicity of notation), the average number of unfolded units $\langle n_u\rangle$, and the average number  of interfaces $\langle \iota\rangle$ between folded and unfolded units. Since 
\begin{equation} \label{interfaces}
\sum_{i=1}^{N-1}S_{i}S_{i+1}=N-1-2\iota,
\end{equation} 
using the dimensionless parameters introduced before, we get
\begin{align}
\langle \tilde{x} \rangle&=\frac{1}{\tilde{\beta}}\frac{\partial\log Z_G(\tilde{f})}{\partial \tilde{f}},\label{<xn>_adim}\\
\langle n_u\rangle&=-\frac{1}{\tilde{\beta}}\frac{\partial \log Z_G(\tilde{f})}{\partial \Delta \tilde{E}},\label{<u>_adim}\\
\langle \iota\rangle&=\frac{N-1}{2}-\frac{1}{2\tilde{\beta}}\frac{\partial\log Z_G(\tilde{f})}{\partial \varpi},\label{<i>_adim}
\end{align}
where $\tilde{x}=x/\ell$ and $\Delta \tilde{E}=\Delta E/J$.
In Fig.\ref{fig:ising1DdiffwellG}, the behavior of the mechanical quantities defined in Eqs.\eqref{<xn>_adim}, \eqref{<u>_adim}, and \eqref{<i>_adim} is shown by varying the values of $\tilde{\beta}$ (different colors) and $\alpha$ (different rows).
More precisely, each row of Fig.\ref{fig:ising1DdiffwellG} corresponds to different values of $\alpha$ (namely, $1/3$, $1$ and $3$ in the first, second and third row, respectively), while $\tilde{\beta}=J/K_BT$ varies in each single plot (assuming the values $1/2$, $1$ and $3/2$ for the blue, yellow and red curves, respectively).
The first column shows the average chain length $\langle \tilde{x} \rangle$, the second one the average number of unfolded units $\langle n_u \rangle$, and the third one the average number of interfaces $\langle \iota\rangle$.

%
%----------------------------------------------------

To begin, considering the force-extension figure in the case with identical wells $\alpha=1$, we can notice that at the higher dimensionless Ising parameter $\tilde{\beta}=3/2$ (red curve) there is a force plateau that corresponds to a cooperative unfolding of the chain units, a typical behavior characterizing the Gibbs ensemble \cite{p1}. 
At the lower value $\tilde{\beta}=1/2$ (blue curve), when interaction terms decrease, this plateau is less sharp due to the fact that thermal fluctuations induce the system to explore a larger part of the wiggly energy landscape, eventually smoothing the force plateau.
Again for $\alpha=1$, the cooperative behavior is further confirmed by the expectation value of the number of unfolded units $\langle n_u\rangle$ that shows a transition from $0$ to $N$ at the same threshold of the force-extension curve.
In the plot of the average number of interfaces $\langle \iota \rangle$ (always with $\alpha=1$), we can notice that increasing the  parameter $\tilde{\beta}$ decreases the number of interfaces due to the fact that in a ferromagnetic scenario all the units tend to be in the same configuration (either folded or unfolded).
Furthermore, in this plot, we can easily notice that the force thresholds that are responsible of the synchronized unfolding of the chain units are the same for different values of $\tilde{\beta}$.
This is due to the fact that in the second row we have considered chain units that present the same elastic constant for both wells ($\alpha=1$).
Consider now the first row ($\alpha=1/3$), when the second well is more compliant (softening regime). In this case, the fully unfolded configuration force-extension curve has a lower slope than the homogeneous folded one.
The most interesting aspect here, is that the force plateau occurs at different thresholds depending on the value of $\tilde{\beta}$ and, then, on the temperature $T$.
This behavior is further confirmed by the shift of the peak in the average number of interfaces $\langle \iota \rangle$, corresponding to a force threshold depending on the value of $\tilde{\beta}$.
In this plot, in fact, we can observe that the force threshold increases with $\tilde{\beta}$, meaning that the force plateau decreases with the temperature.
All these consideration holds even in the third row ($\alpha=3$), when the second state is stiffer (hardening regime), with the only difference that now the force plateau increases with the temperature.

While these results, representing the main physical effects of the proposed model, can already be qualitatively discussed, as described above, in the limiting cases of $N\rightarrow\infty$ and $\tilde{\beta}\gg 1$ this dependence can be quantitatively described, so that a more detailed analysis is postponed later.

%----------------------------------------------------------------------------
\subsection{Ising model within the Helmholtz ensemble}
\label{Helmholtz}

Consider now the isometric loading condition, described by the Helmholtz ensemble (see Fig.\ref{fig:BSPE}, bottom panel). In this case, the total elongation of the chain is fixed by assigning $x_N$.
As shown in Ref.\cite{aes}, one may use an inverse Laplace transform to obtain the canonical partition function in the Helmholtz ensemble, starting from the Gibbs one given in Eq.\eqref{dimension_Z_G}. Using the change of variable $f\to -i\omega/\beta$, we can write
\begin{eqnarray}
Z_H(x_N)=\frac{1}{2\pi}\int_{-\infty}^{+\infty}Z_G\left(-\frac{i\omega}{\beta}\right)e^{i\omega x_N } \mathrm{d}\omega,
\label{Z_H_Lagrange1}
\end{eqnarray}
and we obtain (see Appendix~\ref{appB} for the detailed calculation)
\begin{equation}
Z_H(\tilde{x}_N)=  \frac{\ell^{N-1} e^{N\tilde{\beta}}}{2\cosh \tilde{\beta}}\left(\frac{\pi}{2\tilde{\beta}}\right)^{\frac{N-1}{2}} \left\lbrace\sum_{k=0}^{\left[\frac{N}{2}\right]}\binom{N}{2k} \mathcal{W}_k +e^{-2\tilde{\beta}}\sum_{k=0}^{\left[\frac{N-1}{2}\right]}\binom{N}{2k+1}\mathcal{W}_k\right\rbrace,
\label{zhie}
\end{equation}
where
\begin{equation}
\begin{split}
\mathcal{W}_k= 	&\sum_{j=0}^k\sum_{s=0}^{N-2j}\binom{k}{j}\binom{N-2j}{s}e^{-(s+j)\tilde{\beta} \Delta \tilde{E}}\sqrt{\frac{1}{\tilde{k}^{N-s-j}(\alpha\tilde{k})^{s+j}}}\\
			&\times	(-1)^j4^{j}\left(1-e^{-4\tilde{\beta}}\right)^j\sqrt{\frac{1}{\left( \frac{N-s-j}{\tilde{k}}+\frac{s+j}{(\alpha\tilde{k})}\right)}}\\
			&\times	\exp\left\lbrace-\frac{\tilde{\beta}}{2}\frac{\left[\tilde{x}_N-(N-s-j+\chi s+\chi j)\right]^2}{\left(\frac{N-s-j}{\tilde{k}}+\frac{s+j}{(\alpha\tilde{k})}\right) }\right\rbrace.
\end{split}
\label{wjs}
\end{equation}
%----------------------------------------------------
%
\begin{figure}[t]
    \centering
    \includegraphics[scale=1]{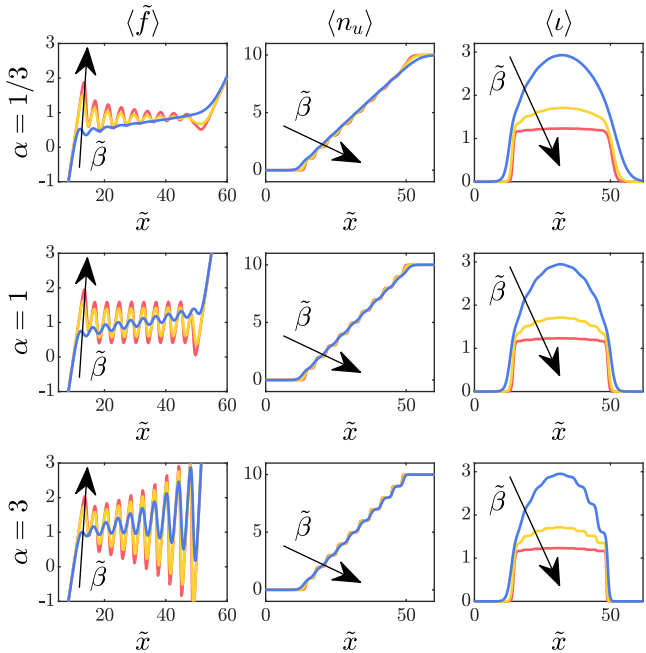}
    \caption{Behavior of the non-convex non-local chain within the Helmholtz ensemble with variable Ising coefficient $\tilde{\beta}$ and parameter $\alpha$. The quantities $\langle \tilde{f}\rangle$, $\langle n_u\rangle$ and $\langle \iota\rangle$ are represented versus the dimensionless extension $\tilde{x}$, with variable $\tilde{\beta}=\{1/2, 1, 3/2\}$. In the first row we used $\alpha=1/3$, in the second one $\alpha=1$, and in the third one $\alpha=3$. Here $N=10$, $\tilde{k}=6$, $Q(-1)=0$, $Q(+1)=\Delta \tilde{E}=4$, $\chi=5$.}
    \label{fig:ising1DdiffwellGGHH2}
\end{figure} 
%
%----------------------------------------------------
This partition function allows us to evaluate the expectation values $\langle f\rangle$ of the force conjugated to the assigned displacement, the average value of the number of unfolded units $\langle n_u\rangle$, and the expectation value of the number of interfaces $\langle \iota\rangle$. We have 
\begin{align}
\langle \tilde{f}\rangle=&-\frac{1}{\tilde{\beta}}\frac{\partial\log Z_H(\tilde{x}_N)}{\partial \tilde{x}_N},\label{xn_helmholtz_adim}\\
\langle n_u\rangle=&-\frac{1}{\tilde{\beta}}\frac{\partial \log Z_H(\tilde{x}_N)}{\partial \Delta \tilde{E}},\label{u_helmholtz_adim}\\
\langle \iota\rangle=&\frac{N-1}{2}-\frac{1}{2\tilde{\beta}}\frac{\partial\log Z_H(\tilde{x}_N)}{\partial \varpi}.\label{i_helmholtz_adim}
\end{align}

In Fig.\ref{fig:ising1DdiffwellGGHH2}, we show the behavior of the system within the Helmholtz ensemble.
As before, we have different values for $\alpha$ in different rows (namely, $\alpha=1/3,\,1\,,3$ in the first, second and third row, respectively) and, in each single plot, $\tilde{\beta}$ is variable (namely, $\tilde{\beta}=1/2, 1, 3/2$ for the blue, yellow and red curves, respectively).
The average values $\langle \tilde{f}\rangle$, $\langle n_u\rangle$ and $\langle \iota \rangle$ are represented versus $\tilde{x}$ in the first, second and third column of Fig.\ref{fig:ising1DdiffwellGGHH2}.
We can observe that the expectation values of the normalized force versus the applied extension always exhibits the typical saw-tooth path associated to a non-synchronized phase transition of the units.
In the Helmholtz ensemble, in fact, due to the prescribed total elongation of the chain, we observe a sequential unfolding rather than a synchronized unfolding of the units, as seen previously within the Gibbs ensemble.
In the plots of the average number of interfaces (third column), we can notice that at the highest $\tilde{\beta}$ value, there is only one single domain wall throughout all the unfolding process.
This means that with high values of the Ising parameter (or similarly for lower value of the temperature) it is strongly disadvantaged to create more than one interface in the course of the sequential unfolding process.
The important result to underline, as already shown in Ref.\cite{aes}, is that when a large ferromagnetic Ising coefficient is considered ($\tilde{\beta}\gg 1$), the system favors the generation of a single  propagating interface, with a nucleation stress peak corresponding to the sudden transition of a chain fraction in the unfolded conformation.
The asymmetry of the curves representing the number of interfaces and the dependence of the force plateau on the temperature (through $\tilde{\beta}$) are related once again to the presence of different elastic constants for the folded and unfolded wells (for the cases with $\alpha\neq 1$).
We remark that from a theoretical point of view, the unfolding of the last unit is characterized by a downward force peak that represents the coalescence of the folded phase to the unfolded one.
Actually, in real experiments this down peak is typically not attained due to the presence of grips forbidding full propagation of the new phase in the terminal region. 
Moreover, when $\tilde{\beta}$ is low (meaning that the thermal fluctuations are high compared to the Ising parameter $J$), we can observe a reduction of the force peaks resulting in a smoothing of the force-extension curve, as shown in the first column of the figure.
The behavior of the first peak is largely analyzed in the following Sections.
The steps observed in the curves representing the number of unfolded units  $\langle n_u\rangle$ versus the rising normalized extension are also smoothed with a decreasing $\tilde{\beta}$. Similarly, the number of interfaces $\langle \iota\rangle$ increases with lower values of $\tilde{\beta}$.

%---------------------------------------------------------------------------------------------------------------------------------------
\section{Thermodynamic limit with strongly ferromagnetic behavior in the Gibbs ensemble}
\label{TL-Gibbs}

As anticipated, the plateau force observed in the force-extension curves, corresponding to the so called Maxwell stress in the purely mechanical case \cite{pt1}, sensibly depends on $\tilde{\beta}$ and, therefore, on temperature in both the cases with applied force (Gibbs) and with total fixed elongation (Helmholtz).
Here, to obtain an analytical measure of such an important effect, we consider strongly Ising interactions, \textit{i.e.} with $\tilde{\beta}=\frac{J}{k_B T}$ sufficiently large, and we study the system in the thermodynamic limit ($N\to\infty$). An analogous limit was considered in Ref.\cite{p2} for a purely mechanical system. We remark that this limit, due to the complexity of the system, is performed here only for the Gibbs ensemble. In the following Sections, we will introduce particular hypothesis to get more analytical results also in the Helmholtz ensemble. 

Let us consider first the thermodynamic limit for the Gibbs ensemble. Thus, from Eq.\eqref{ZGwithlambdas} we have 
\begin{equation}
\log Z_G \underset{N\to\infty}{\sim}N\log{\hat \lambda_1},
\end{equation}
where we exploited the property ${\hat \lambda_1}>{\hat \lambda_2}$ (see Eq.~(\ref{eigvalT}) for details). From Eq.(\ref{<xn>_adim}) we find
\begin{equation}
\langle \tilde{x}\rangle \simeq \frac{N}{\tilde{\beta}}\frac{\partial}{\partial \tilde{f}}\log{\hat \lambda_1}.
\label{xaspett}
\end{equation}
A direct evaluation of the derivative in Eq.~\eqref{xaspett} gives 
\begin{equation}
\frac{\langle \tilde{x}\rangle}{N}\simeq\frac{1}{2}\left[\left(1+\frac{c_--c_+}{\sqrt{\Delta}}\right)\left(1+\frac{\tilde{f}}{\tilde{k}}\right)+\left(1-\frac{c_--c_+}{\sqrt{\Delta}}\right)\left(\chi+\frac{\tilde{f}}{\alpha\tilde{k}}\right)\right],
\label{F-Xwithlambdas}
\end{equation}
where $c_-$ and $c_+$ also depend on $\tilde f$ and are defined in Eq.~\eqref{cpm} and $\Delta$ in Eq.~(\ref{delta}) (see Appendix~\ref{appB}).
In Fig.\ref{fig:xLTD.eps}, we show an example of application of Eq.\eqref{F-Xwithlambdas}, where some force-extension curves are plotted with different values of $\tilde{\beta}$ (left panel) and $\alpha$ (right panel).
Specifically, on the left panel of Fig.\ref{fig:xLTD.eps}, we can observe that increasing $\tilde{\beta}$ has both the effect of reducing the slope of the force plateau and of increasing the Maxwell force, confirming once again that the force plateau depends on the value of  $\tilde{\beta}$ and then on the temperature $T$. Observe that here we are considering a fixed value of $\alpha=\frac{1}{3}$ with the second phase softer than the first one.
On the right panel of Fig.\ref{fig:xLTD.eps}, instead, we can see that keeping $\tilde{\beta}$ constant and increasing $\alpha$ we induce an increase in the force plateau, proving that the Maxwell force strictly depends on the asymmetry of the elastic constants of the two energy wells, as already observed in the purely mechanical case when constant force plateaux depending on $\alpha$ are observed.
%----------------------------------------------------
%
\begin{figure}[t]
    \centering
    \includegraphics[scale=0.9]{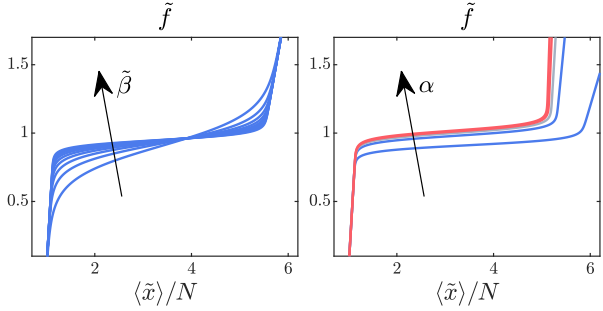}
    \caption{Force-extension curves within the Gibbs ensemble in the thermodynamic limit. In the left panel, $\tilde{\beta}=\{2, 3, \dots, 10\}$ is varied while $\alpha=1/3$ is constant. On the right panel, instead, $\tilde{\beta}=10$ is constant while $\alpha=\{0.2, 0.6, 1, 1.4, 1.8\}$ is variable. The curves are blue for $\alpha<1$ and red for $\alpha>1$ (the grey curve corresponds to $\alpha=1$). The other parameters are $\tilde{k}=6$, $\tilde{Q}(-1)=0$, $\tilde{Q}(+1)=\Delta \tilde{E}=4$, $\chi=5$, $N=10$.
    }
    \label{fig:xLTD.eps}
\end{figure}
%
%----------------------------------------------------

Interestingly, it is possible to link the expectation value of the number of unfolded units to the assigned force in the thermodynamic limit.
Indeed, by using Eq.\eqref{<u>_adim} for $N\to\infty$, we get 
\begin{equation}
\begin{array}{l} \displaystyle
\frac{\langle n_u\rangle}{N} \simeq -\frac{1}{\tilde{\beta}}\frac{\partial}{\partial \Delta \tilde{E}} \log{\hat \lambda_1}=\frac{1}{2} \left(1-\frac{c_--c_+}{\sqrt{\Delta}}\right), \vspace{0.2 cm}\\
\displaystyle \frac{\langle n_f\rangle}{N} = 1- \frac{\langle n_u\rangle}{N}  =\frac{1}{2} \left(1+\frac{c_--c_+}{\sqrt{\Delta}}\right).\end{array}
\label{numbers}
\end{equation}
where again $c_-$ and $c_+$ depend on $\tilde f$ and are defined in Eq.~\eqref{cpm}.
 Thus we may write Eq.\eqref{F-Xwithlambdas} in the form 
\begin{equation}
\label{F-Xwithl}
\frac{\langle \tilde{x}\rangle}{N}\simeq\frac{\langle n_f\rangle}{N} \left(1+\frac{\tilde{f}}{\tilde{k}}\right)+\frac{\langle n_u\rangle}{N}\left(\chi+\frac{\tilde{f}}{\alpha\tilde{k}}\right).
\end{equation}
We notice that Eq.~\eqref{F-Xwithl}, derived from Eqs.\eqref{F-Xwithlambdas} and \eqref{numbers}, gives a direct physical interpretation of the transition process with the expected value of the elongation obtained as a convex combination of the stretch-force relations in the purely folded and purely unfolded phases, depending on their percentage. In particular, when $\langle n_u\rangle=0$, all units are in the folded phase and the system follows the first branch as obtained without temperature and non local interaction effects. On the other hand, in the opposite extreme case with $\langle n_u\rangle=N$, all the bistable units are unfolded and the elongation is now given by  $\langle \tilde{x}\rangle=\chi N+ (N/\alpha \tilde{k})\tilde{f}$. These two homogeneous regimes, where the energy is convex and temperature has no effects, are connected by a force plateau whose height and slope depend on both temperature and Ising coefficient $J$ through $\tilde{\beta}$.

\begin{figure*}[t!]
        \includegraphics[scale=0.68]{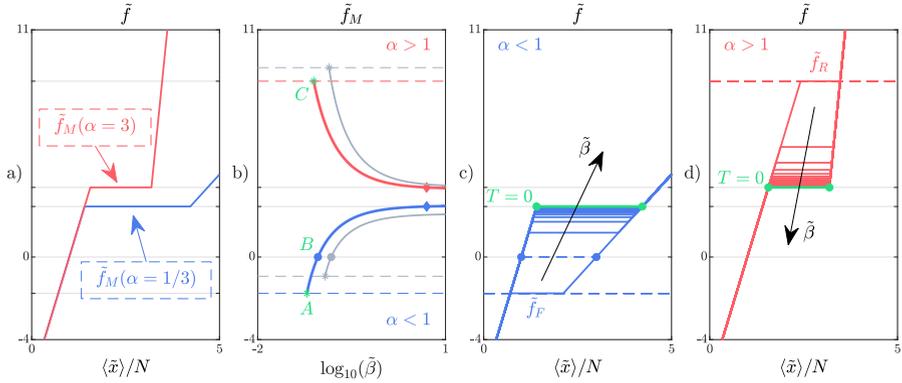}
    \caption{\label{fig:MFOTh<>k} Dependence of the Maxwell force by $\tilde{\beta}$. Panel a): piecewise linear force-extension response in the thermodynamic limit under strong ferromagnetic assumption $\tilde{\beta} \gg 1$. We used $\alpha=1/3$ (solid blue line) and $\alpha=3$ (solid red line). Moreover, we adopted $\tilde{\beta}=5$ for both curves.
Panel b): dimensionless Maxwell force versus $\beta$ with $\alpha=1/3$ and $\alpha=3$ (solid blue and solid red curves, respectively). Also the curves for $\alpha=1/6$ and $\alpha=6$ are shown for completeness (solid grey curves). The theoretical maximum (roof) and minimum (floor) values of the Maxwell forces  $\tilde{f}_{M}(\tilde{\beta}^*)=\tilde f_R$ and $\tilde{f}_{M}(\tilde{\beta}^*)=\tilde f_F$ are represented by dashed blue ($\alpha<1$) and red ($\alpha>1$)  curves. Panels c) and d): evolution of the force plateau in the force-extension curve as $\tilde{\beta}$ changes for the two cases of $\alpha<1$ (panel c) and $\alpha>1$ (panel d). Green plateaux correspond to $T=0$ (see Fig.\ref{fig:wells}).  Adopted parameters: $\tilde{k}=6$, $\tilde{Q}(-1)=0$, $\tilde{Q}(+1)=\Delta \tilde{E}=6$, $\chi=3$.}
    \end{figure*}

Let us now introduce the hypothesis of strong ferromagnetic interactions with respect to temperature, $\tilde{\beta} \gg 1$.
More precisely, from the definition of $\Delta$ in Eq.~(\ref{delta}), it is possible to consider the approximation 
\begin{equation}
\Delta \simeq \left(c_--c_+\right)^2,
\end{equation}
valid if
\begin{equation}\label{eq:ratio}
\frac{4c_+c_- e^{-4\tilde{\beta}}}{(c_--c_+)^2}\ll 1.
\end{equation}
Taking into account the form of $c_{\pm}$ (see Eqs.~\eqref{cs} and \eqref{cpm} in Appendix B), this condition corresponds to 
\begin{equation}
\tilde{\beta}\gg \frac{1}{4}\log\left[\frac{2}{\cosh \left(\delta\right) -1}\right],
\end{equation}
where
\begin{equation}
\label{eq:delta}
\delta=\frac{1}{2}\log\left(\frac{1}{\alpha}\right)-\tilde{\beta}\left[\Delta \tilde{E}- (\chi-1)\tilde{f}-\left(\frac{1}{\alpha}-1\right)\frac{\tilde{f}^2}{2\tilde{k}}\right].
\end{equation}
In particular, one verifies that in the limit $\tilde{\beta}\rightarrow+\infty$ (with $\delta\ne 0$) the ratio in Eq. (\ref{eq:ratio}) goes to zero.
Using this approximation, Eq.~\eqref{F-Xwithlambdas} shows the two following different regimes, corresponding to the two homogeneous folded and unfolded fractions, depending on the force $\tilde f$ through the parameters $c_-=c_-(\tilde f)$ and  $c_+= c_+(\tilde f)$ 
\begin{equation}
    \frac{\langle \tilde{x}\rangle}{N}\simeq
    \left\{
    \begin{aligned}
        &1+\frac{\tilde{f}}{\tilde{k}} &\quad\mbox{if }\quad c_->c_+,\\
        &\chi+\frac{\tilde{f}}{\alpha\tilde{k}} &\quad\mbox{if } \quad c_-<c_+.\\
    \end{aligned}
    \right.
    \label{eq:elasticbranches_}
\end{equation}
Thus under the assumption of large $\tilde{\beta}$, we get a horizontal force plateau, as shown in Fig.\ref{fig:MFOTh<>k} (panel a), obtained by imposing $c_-=c_+$, or, by using Eq.~\eqref{cs}, given by the condition
\begin{equation}
\delta=0,
\label{MF}
\end{equation}
where $\delta$ has been defined in Eq.(\ref{eq:delta}). In this way we obtain the main result of the paper, {\it i.e.} that for large enough interface energy $J$ and for a large value of $N$, the system is characterized by a {\it temperature dependent transition force}. This is controlled by a new temperature dependent term in  Eq.\eqref{MF}, not present in Eq.(\ref{zerotemp}).
Moreover, we can obtain an explicit analytic expression of the dimensionless Maxwell force $\tilde{f}_M$ as follows  
\begin{equation}
\tilde{f}_M(\tilde{\beta})=\left(\sqrt{\mathcal{D}}-1\right) \left(\frac{\alpha}{1-\alpha}\right)\tilde{k}\left(\chi-1\right),
\label{solMF}
\end{equation}
where
\begin{equation}
\mathcal{D}=1-\frac{2}{\tilde{\beta} (\chi-1)^2\tilde{k}}\left(\frac{1-\alpha}{\alpha}\right)\left(\frac{1}{2}\log\frac{1}{\alpha}-\tilde{\beta}\Delta \tilde{E}\right).
\label{ddd}
\end{equation}
Due to the different elastic constants ($\alpha\neq 1$), the Maxwell force obtained from Eq.~\eqref{solMF} depends on the temperature $T$ (through $\tilde{\beta}$).
A simpler dependence can be obtained by expanding Eq.(\ref{solMF}) up to the first order in the difference $(1-\alpha)/\alpha$, measuring the difference from the case when the two wells are identical $\alpha=1$. We obtain 
\begin{equation}
\tilde{f}_M(\tilde{\beta})\simeq\frac{\Delta \tilde{E}}{\chi-1}-\left[\frac{\Delta\tilde{E}^2}{2\tilde{k}(\chi-1)^3}+\frac{1}{2\tilde{\beta}(\chi-1)}\right]\left(\frac{1-\alpha}{\alpha}\right),
\label{solMFapp}
\end{equation}
which is valid for $\mid(1-\alpha)/\alpha\mid\ll1$.
When $\alpha=1$ we retrieve the expression $\tilde{f}=\Delta \tilde{E}/(\chi-1)$, well known in literature \cite{soft,JCPnew,bell1,bell2}, and discussed in the Introduction. On the other hand, if $T$ approaches zero, Eq.(\ref{MF}) simplifies to Eq.(\ref{zerotemp}).

The dependence of the Maxwell force on $\tilde{\beta}$ is shown in Fig.\ref{fig:MFOTh<>k} (panel b). We can observe the opposite behaviors exhibited in the hardening case with $\alpha>1$  (solid blue curve with $\alpha=1/3$) and in the softening case with $\alpha<1$ (solid red curve with $\alpha=3$). We can also notice that when $\tilde{\beta}\to\infty$ the Maxwell force reads
\begin{equation}
\tilde{f}_{M} \simeq\tilde{k}(\chi-1)\frac{\alpha}{1-\alpha}\Biggl[\sqrt{1+\frac{2\Delta \tilde{E}}{(\chi-1)^2 \tilde{k}}\left(\frac{1-\alpha}{\alpha}\right)}-1\Biggr]>0,
\label{eq:maxf0}
\end{equation}
with a first order correction $O(1/\tilde{\beta})$. In particular, when $\alpha < 1$, if $\tilde{\beta}$ decreases (as the temperature increases), the Maxwell force decreases as well. The corresponding variation in the stress-strain diagram is shown in Fig.\ref{fig:MFOTh<>k} (panel c). 
In particular, from Eqs.(\ref{solMF}) and (\ref{ddd}), we deduce that when the value of $\tilde{\beta}$ satisfies the equality $\log(1/\alpha)/2\Delta \tilde{E}=\tilde{\beta}$, the Maxwell force characterizing the plateau is $\tilde{f}_{M}=0$ (see the point $B$ in  Fig.\ref{fig:MFOTh<>k}, panel b). Thus, at this value $\tilde{\beta}=\tilde{\beta}_0$, the Maxwell plateau at zero applied force connects the two natural configurations with $\langle \tilde{x}\rangle/N\simeq1$ and $\langle \tilde{x}\rangle/N\simeq\chi$ corresponding to the completely folded and completely unfolded states, respectively. 
In other word, the system undergoes a transition to the second state at zero applied force. It is important to observe that solutions with zero or even negative Maxwell forces have been considered in the case of phase transitions in shape memory alloys \cite{Seelecke}. In this case a consistent definition of the reference configuration, assumed to be stable, should be temperature dependent, so that for positive (negative) Maxwell stress the reference configuration is the homogeneous folded (unfolded) phase configuration.  

It is important to observe that, for this case with $\alpha<1$, the Maxwell stress can decrease until the limit value $\tilde f_F$, as defined in Eq.\eqref{fF}, because after this value the energy of the unfolded state is not defined (see the point A in Fig.\ref{fig:MFOTh<>k}, panel b, and the floor plateau in Fig.\ref{fig:MFOTh<>k}, panel c).  We denote $\tilde{\beta}^*$ the value of $\tilde{\beta}$, for which we attain the plateau at $\tilde f_F$. Interestingly, for $\alpha<1$, the equation $\tilde{f}_{M}(\tilde{\beta}^*)=\tilde f_F$ is solved by
\begin{eqnarray}
\tilde{\beta}^*=\frac{\frac{1}{2}\log\frac{1}{\alpha}}{\Delta \tilde E (1-\alpha)+\frac{1}{2}\alpha\tilde k (1-\chi)^2}.
\end{eqnarray}

In the case with $\alpha > 1$, represented in Fig.\ref{fig:MFOTh<>k}, panel b (red solid curve), we observe that the Maxwell stress always increases with the temperature (decreasing $\tilde{\beta}$). In this situation, the Maxwell force is always positive. However, the plateau stops existing at $\tilde{\beta}^*$ corresponding to the value of force $\tilde f_R$, defined in Eq.\eqref{fR}, when the energy of the first well ceases to be defined (see the point C in Fig.\ref{fig:MFOTh<>k}, panel b, and the roof plateau in Fig.\ref{fig:MFOTh<>k}, panel d). Interestingly, for $\alpha>1$, the equation $\tilde{f}_{M}(\tilde{\beta}^*)=\tilde f_R$ is solved by
\begin{eqnarray}
\tilde{\beta}^*=\frac{\frac{1}{2}\alpha\log\alpha}{\Delta \tilde E (1-\alpha)+\frac{1}{2}\alpha\tilde k(1-\chi)^2}.
\end{eqnarray}
It is important to point out that, due to the use of the spin variables method, the temperature is always limited by the fact that the barrier between the two energy wells must always be sufficiently larger than $K_BT$.

In this Section, we obtained an analytical expression ($\delta=0$) linking the Maxwell force to the temperature in the thermodynamic limit ($N\to\infty$) and with the assumption of strong ferromagnetic interaction (under the isotensional condition). This result is related to the difference of stiffness between the two energy wells ($\alpha\neq1$) and explains the mechanical response observed in several nanosystems \cite{dmo0,dmo0b,dmo0c,dmo1,dmo2,dmo3,dmo3b,dmo4,dmo5,dmo6,dmo7,dmo8,dmo9,dmo10}.  

Interestingly enough, if for $\alpha >1$ ($\alpha <1$) we increase (decrease) the stress starting from the homogeneous folded (unfolded) state, the system keeps this configuration even for values of the stress for which this configuration is characterized by a higher elastic energy than the other homogeneous state, due to entropic effects. This counter-intuitive behavior, tending to stabilize the softer phase, is observable only in the case of different wells and  was named entropic stabilization \cite{Seelecke}.

%---------------------------------------------------------------------------------------------------------------------------------------
\section{The zipper model}
\label{Zipper}

\begin{figure}[t]
    \centering
    {\includegraphics[scale=0.8]{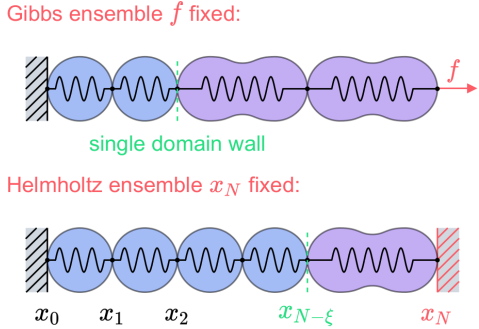}}
    \caption{Scheme of the zipper model of the non-local non-convex chain. Top panel: isotensional configuration (Gibbs ensemble). Bottom panel: isometric configuration (Helmholtz ensemble). In both case we always have only one interface between folded and unfolded regions. }
    \label{fig:zipper}
\end{figure}

In this Section, we further extend the study of the non-local non-convex chain of bistable units under the assumption of strong ferromagnetic behavior ($\tilde{\beta}\gg 1$).
As previously discussed, this hypothesis leads to the existence of a single domain wall \cite{p2,aes}, \textit{i.e} a single interface between the folded and unfolded regions that propagates continuously through the chain (see Fig.\ref{fig:zipper}).
In this framework, we derive simplified analytical expressions for both the Gibbs and Helmholtz ensembles and we prove their equivalence in the thermodynamic limit.
Accordingly, we consider a chain composed by $N-\xi$ units in the folded state and the remaining $\xi$ units in the unfolded one, being $\xi$ a discrete variable assuming values in the set $\{0,1,2,\dots,N\}$, representing the position, in terms of chain units, of the moving domain wall between folded and unfolded regions.
This discrete variable can vary depending on both mechanical and thermal effects and this simplified scheme is typically called zipper model~\cite{zip1,zip2,zip3,zip4}.

To begin with, let us consider the Hamiltonian in Eq.\eqref{overallHa}, where the last sum over the $N-1$ spins can be divided into two parts by means of the zipper assumptions.
By introducing $\iota$, the number of changes (interfaces) in the spins sequence $S_1,\dots,S_N$, we have $\iota$ addends with value $-1$ and $N-1-\iota$ addends with value $+1$, see Eq.(\ref{interfaces}).
Thus 
\begin{equation}
\label{H_Z_I}
\tilde{H}_Z = \sum_{i=1}^{N}\left\lbrace \tilde{Q}(S_i)+\frac{\tilde{K}(S_i)}{2}\left(\lambda_i-\lambda_0(S_{i})\right)^2\right\rbrace-[N-1-2\iota(\xi)].
\end{equation}
In particular, we assume the possibility to have $\iota(\xi)=0$ if $\xi\in\{0,\,N\}$ (no interface) and $\iota(\xi)=1$ if $1\leq\xi\leq N-1$ (only one interface). Accordingly: (i) if $i\leq N-\xi$ (folded units), then we have $\tilde{Q}(i)=0$, $\tilde{K}(i)=\tilde{k}$, $\lambda_0(i)=1$, and (ii) if $i\geq N-\xi+1$ (unfolded units), then $\tilde{Q}(i)=\Delta \tilde{E}$, $\tilde{K}(i)=\alpha \tilde{k}$, $\lambda_0(i)=\chi$.

By separating the folded and unfolded contributions we may rearrange the Hamiltonian as 
\begin{equation}
\tilde{H}_Z(\xi)=\sum_{i=1}^{N-\xi}\left\lbrace\frac{\tilde{k}}{2}\left(\lambda_i-1\right)^2\right\rbrace+ \sum_{i=N-\xi+1}^{N}\left\lbrace \Delta \tilde{E}+\frac{\alpha\tilde{k}}{2}\left(\lambda_i-\chi\right)^2\right\rbrace -\left[N-1-2\iota(\xi)\right].
\label{Hzipper}
\end{equation}
From now on, the discrete variable $\xi$ belongs to the phase space of the system together with the continuous displacements $\lambda_i$.
%----------------------------------------------------------------------------
\subsection{Zipper model within the Gibbs ensemble}
\begin{figure}[t]
    \centering
    \includegraphics[scale=1]{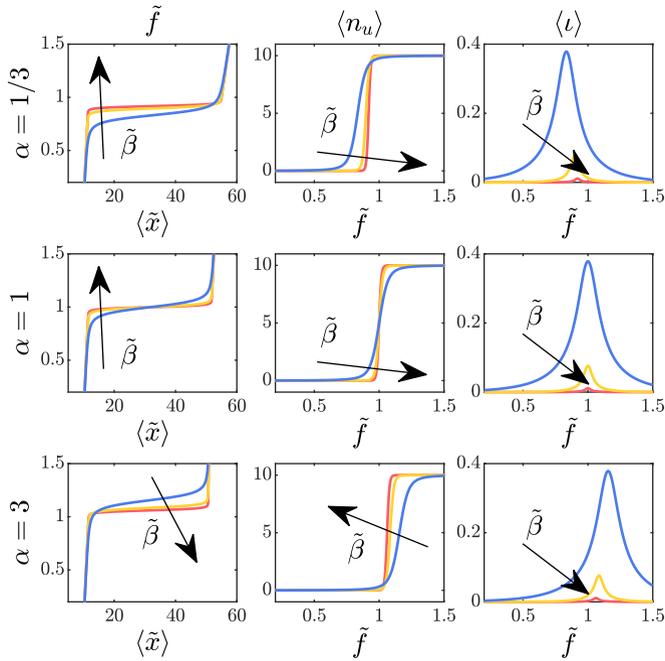}
    \caption{Behavior of the zipper model within the Gibbs ensemble with variable Ising coefficient $\tilde{\beta}$ and parameter $\alpha$. The quantities $\langle \tilde{x}\rangle$, $\langle n_u\rangle$ and $\langle \iota\rangle$ are represented versus the dimensionless force $\tilde{f}$, with a variable Ising coefficient $\tilde{\beta}=\{1, 2, 3\}$. In the first row we used $\alpha=1/3$, in the second one $\alpha=1$, and in the third one $\alpha=3$. We adopted the parameters $N=10$, $\tilde{k}=6$, $Q(-1)=0$, $Q(+1)=\Delta \tilde{E}=4$, $\chi=5$.}
    \label{fig:isingzipGH_Gibbs}
\end{figure}

Let us evaluate the canonical partition function in the Gibbs ensemble by using the Hamiltonian in Eq.~\eqref{Hzipper}. By definition we have 
\begin{equation}
Z_G(\tilde{f})=\ell^N\sum_{\xi=0}^{N}\int_{\mathbb{R}^{N}} e^{-\tilde{\beta} \left(\tilde{H}_Z- \tilde{f}(\sum_{i=1}^N\lambda_i)\right)} \mathrm{d}\lambda_1 \dots\mathrm{d}\lambda_N,
\end{equation}
that can be evaluated by a Gaussian integration giving
\begin{equation}
    Z_G(\tilde{f})	=\ell^N\left(\frac{2\pi}{\tilde{\beta}\tilde{k}}\right)^\frac{N}{2}e^{\tilde{\beta} (N-1)}\sum_{\xi=0}^N\frac{1}{\alpha^{\frac{\xi}{2}}}e^{-\tilde{\beta}\left\lbrace 2\iota(\xi)+\Delta \tilde{E} \xi -\tilde{f}[N+(\chi-1)\xi]-\frac{\tilde{f}^2}{2\tilde{k}}\left[N-\xi+\frac{\xi}{\alpha}\right]\right\rbrace}.
\label{Z_G_I}
\end{equation}
Following the same reasoning as in Section~\ref{Isingmodel}, one may evaluate the expectation values of the mechanical macroscopic observables of the system, namely $\langle \tilde{x}\rangle$, $\langle n_u\rangle$ and $\langle \iota\rangle$ (see Eqs.\eqref{<xn>_adim}, \eqref{<u>_adim} and \eqref{<i>_adim}), where we use again $\tilde{x}$ to indicate $\tilde{x}_N$.
In Fig.\ref{fig:isingzipGH_Gibbs}, the behavior of the mechanical quantities $\langle \tilde{x}\rangle$, $\langle n_u\rangle$ and $\langle \iota\rangle$ is shown by varying the values of $\tilde{\beta}$ (different colors) and  $\alpha$  (different rows).
Each row of Fig.\ref{fig:isingzipGH_Gibbs} corresponds to a different constant $\alpha$ (namely, $1/3$, $1$ and $3$ in the first, second and third row, respectively) while $\tilde{\beta}=J/K_BT$ varies within each single plot ($\tilde{\beta}=1, 2, 3$ for the blue, yellow and red curves, respectively).
The first column shows the average chain length $\langle \tilde{x} \rangle$, the second one the average number of unfolded units $\langle n_u \rangle$ and the third one the average number of interfaces $\langle \iota\rangle$.
By looking at the force-extension curve and at the average number of unfolded units, one observes again the typical synchronized behavior of the Gibbs ensemble, where all the units unfold cooperatively.
A difference with respect to the exact Ising model is that the number of interfaces is smaller, limited by the presence of the zipper assumption. 

%----------------------------------------------------
%
%
%----------------------------------------------------

In particular, we may study $\langle x\rangle$, evaluated through Eq.\eqref{<xn>_adim} and the partition function in Eq.\eqref{Z_G_I}. After some straightforward calculations, one obtains
\begin{equation}
\label{x/nl_}
\frac{\langle \tilde{x}\rangle}{N} =
\frac{\left[1+\frac{\tilde{f}}{\tilde{k}}+\left(\chi + \frac{\tilde{f}}{\alpha\tilde{k}}\right)e^{N\delta}\right]\left(1-e^{-2\tilde{\beta}}\right)+e^{-2\tilde{\beta}}\displaystyle\sum_{\xi=0}^{N}e^{\delta \xi}A_\xi}
{\left[1+e^{N\delta}\right]\left(1-e^{-2\tilde{\beta}}\right)+e^{-2\tilde{\beta}}\displaystyle\sum_{\xi=0}^{N}e^{\delta \xi}},
\end{equation}
where $\delta$ is defined in Eq.\eqref{eq:delta} and
\begin{equation}
A_\xi = \frac{\xi}{N}\left(\chi+\frac{\tilde{f}}{\alpha\tilde{k}}\right)+\left(1-\frac{\xi}{N}\right)\left(1+\frac{\tilde{f}}{\tilde{k}}\right).
\end{equation}
We can now use the following sums
\begin{align}
\sum_{\xi=0}^N \mathrm{y}^{\xi}&=\frac{1-\mathrm{y}^{N+1}}{1-\mathrm{y}},\\
\sum_{\xi=0}^N \xi\mathrm{y}^{\xi}&=\frac{\mathrm{y}}{(1-\mathrm{y})^2}[1-(N+1)\mathrm{y}^N+N\mathrm{y}^{N+1}],
\end{align}
which allow us to obtain the explicit expressions
\begin{align}
\nonumber
\Sigma_1=\sum_{\xi=0}^N e^{\delta\xi}A_\xi
=&\left(1+\frac{\tilde{f}}{\tilde{k}}\right)\frac{1-e^{\delta(N+1)}}{1-e^{\delta}}+\left[\left(\chi+\frac{\tilde{f}}{\alpha\tilde{k}}\right)-\left(1+\frac{\tilde{f}}{\tilde{k}}\right)\right]&\\
&\times \frac{1}{N}\frac{e^\delta}{(1-e^\delta)^2}\left[1-(N+1)e^{N\delta}+Ne^{\delta(N+1)}\right],\\
\Sigma_2=\sum_{\xi=0}^N e^{\delta\xi}=&\frac{1-e^{\delta(N+1)}}{1-e^{\delta}}.
\end{align}
Thus, Eq.\eqref{x/nl_} reads
\begin{equation}
\label{end_pos_sigma}
\frac{\langle \tilde{x}\rangle}{N} =
\frac{\left[1+\frac{\tilde{f}}{\tilde{k}}+\left(\chi + \frac{\tilde{f}}{\alpha\tilde{k}}\right)e^{N\delta}\right]\left(1-e^{-2\tilde{\beta}}\right)+e^{-2\tilde{\beta}}\Sigma_1}
{\left[1+e^{N\delta}\right]\left(1-e^{-2\tilde{\beta}}\right)+e^{-2\tilde{\beta}}\Sigma_2},
\end{equation}
which provides the Gibbs force-extension relation in the approximations of single interface and large ferromagnetic interaction (zipper model). Observe that, thanks to this hypothesis, we are able to obtain for the zipper model exact explicit relations for the material response of the system also in the discrete case with finite arbitrary $N$.

On the other hand, in the thermodynamic limit ($N\to\infty$), we can confirm the results obtained in Section \ref{TL-Gibbs} for the complete Ising scheme. In particular, when $N\to\infty$, the value of $\langle \tilde{x}\rangle/N$ depends on the sign of $\delta$. If $\delta<0$ we have
\begin{equation}
\lim_{N\to\infty} \Sigma_1 = \left(1+\frac{\tilde{f}}{\tilde{k}}\right)\frac{1}{1-e^\delta},\,\,\,\,\,\,\,\,\,
\lim_{N\to\infty} \Sigma_2 = \frac{1}{1-e^\delta},
\end{equation}
and we get (in the limit $N\to\infty$)
\begin{equation}
\frac{\langle \tilde{x}\rangle}{N}\simeq1+\frac{\tilde{f}}{\tilde{k}},
\end{equation}
that represents the elastic branch of the response observed when all units are folded. Conversely, if $\delta>0$, we can show that
\begin{equation}
\lim_{N\to\infty} \Sigma_1e^{-N\delta} =-\left(\chi +\frac{\tilde{f}}{\alpha\tilde{k}}\right)\frac{e^\delta}{1-e^\delta},\,\,\,\,\,\,\,\,\,
\lim_{N\to\infty} \Sigma_2e^{-N\delta} = -\frac{e^\delta}{1-e^\delta}.
\end{equation}
Hence, a manipulation of Eq.(\ref{end_pos_sigma}) gives (in the limit $N\to\infty$)
\begin{equation}
\frac{\langle \tilde{x}\rangle}{N}\simeq\chi+\frac{\tilde{f}}{\alpha\tilde{k}},
\end{equation}
representing the unfolded elastic branch. Thus, the two elastic branches for $\delta<0$ and $\delta>0$ are linked by a force plateau (see Fig.\ref{fig:MFOTh<>k}, panel a), corresponding to Eq. $\delta=0$. This is exactly the same condition found in Eq.\eqref{MF}, proving that the results obtained within the zipper approximation in the thermodynamic limit coincide with the ones obtained in Sect.\ref{TL-Gibbs} under the strong ferromagnetic assumption.

%----------------------------------------------------------------------------
\subsection{Zipper model within the Helmholtz ensemble}
\label{Zipper-Helmholtz}

Let us then consider the Helmholtz ensemble under the zipper assumption. In this case, the partition function can be obtained as in Eq.\eqref{Z_H_Lagrange1}, by the application of a Fourier transform. Indeed, using Eq.\eqref{Z_G_I}, we get 
\begin{equation}
\begin{split}
Z_H(\tilde{x}_N) 	=& \left(\frac{2\pi\ell^2  e^{2\tilde{\beta}}}{\tilde{\beta}}\right)^\frac{N-1}{2}\sum_{\xi=0}^N\sqrt{\frac{1}{\left(\frac{N-\xi}{\tilde{k}}+\frac{\xi}{\alpha\tilde{k}}\right)\tilde{k}^{N-\xi}(\alpha\tilde{k})^{\xi}}}\\
&\times e^{-\tilde{\beta}\left(\frac{[\tilde{x}_N-(N+(\chi-1)\xi)]^2}{2}\left(\frac{N-\xi}{\tilde{k}}+\frac{\xi}{\alpha\tilde{k}}\right)^{-1}+2\iota(\xi)+\Delta \tilde{E}\xi\right)}.
\label{Z_H_compact}
\end{split}
\end{equation}
%
%----------------------------------------------------
%
\begin{figure}[t]
    \centering
    \includegraphics[scale=1]{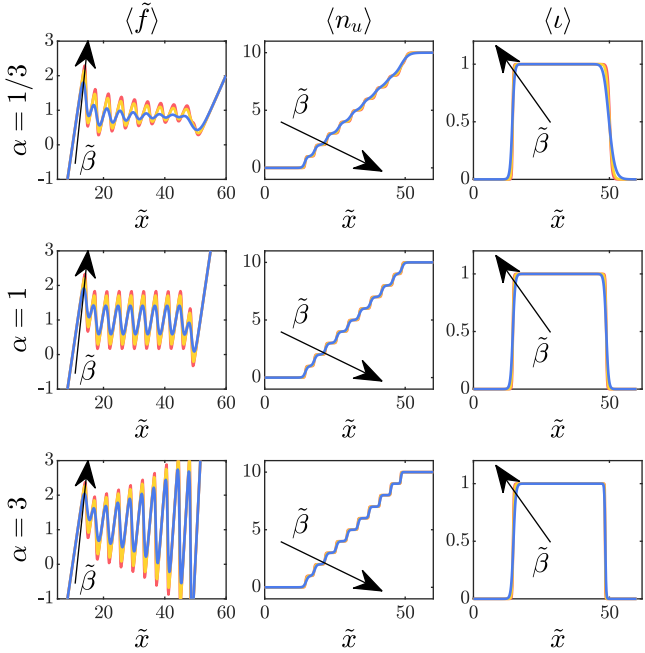}
    \caption{Behavior of the zipper model within the Helmholtz ensemble with variable Ising coefficient $\tilde{\beta}$ and parameter $\alpha$. The quantities $\langle \tilde{f}\rangle$, $\langle n_u\rangle$ and $\langle \iota\rangle$ are represented versus the strain $\tilde{x}$, with a variable Ising coefficient $\tilde{\beta}=\{1, 2, 3\}$. In the first row we used $\alpha=1/3$, in the second one $\alpha=1$, and in the third one $\alpha=3$. We adopted the parameters $N=10$, $\tilde{k}=6$, $Q(-1)=0$, $Q(+1)=\Delta \tilde{E}=4$, $\chi=5$.}
    \label{fig:isingzipGH_Helmholtz}
\end{figure}
%
%----------------------------------------------------
The average macroscopic property of the zipper model within the Helmholtz ensemble can be found through Eqs.\eqref{xn_helmholtz_adim}, \eqref{u_helmholtz_adim} and \eqref{i_helmholtz_adim}. In Fig.\ref{fig:isingzipGH_Helmholtz} we show $\langle \tilde{f}\rangle$, $\langle n_u\rangle$ and $\langle \iota\rangle$ as function of the prescribed extension $\tilde{x}_N$ for different values of the Ising coefficient $\tilde{\beta}$ and different elastic constants ratio $\alpha$. We recover the same macroscopic behavior of the exact Ising model characterized by the sequential unfolding of the units and represented by the typical saw-tooth path of the force-extension curve. In particular, it is important to observe that the value of the number of interface remains constantly at $1$, due to the zipper assumption, and it is $0$ only when $\xi=0$ or $\xi=N$, coherently with the model introduced.
It is interesting to notice that in the force-extension curve, we can see the first upward force peak, corresponding of the nucleation of a new phase with unfolded units, and the last downward peak, representing the coalescence of the folded region to the unfolded one.
Observing the sequential unfolding of the units in the force-extension plots of Fig.\ref{fig:isingzipGH_Helmholtz}, we can notice that there are three different behaviors depending on the elastic constants ratio: in the case with $\alpha=1/3$, the force needed to unfold a single unit decreases while the number of unfolded units increases; in the case with $\alpha=1$, except for the nucleation (first peak) and the coalescence (last peak) forces, all the other forces needed to unfold the units are equal; finally, in the case with $\alpha=3$, we have the opposite behavior to the case with $\alpha=1/3$, \textit{i.e.} the unfolding forces increase with the number of unfolded units.
This result is perfectly coherent with the previously introduced concept of entropic stabilization of the softer configuration. Moreover we remark that by considering higher number of elements (not reported for compactness), it is possible to observe an initial nucleation of a segment with more than one single element passing to the new unfolded state. This aspect is discussed in detail in the next section where we consider the system behavior for large values of $N$. In this case we can analyze the amplitude of the first peak and of the first nucleated segment analytically leading to a useful result for comparing the model with experiments and numerical simulations. 

%---------------------------------------------------------------------------
\section{Stationary phase analysis of the zipper model within the Helmholtz ensemble}

Here, we want to study the thermodynamic limit of the Helmholtz zipper model. The aim of this development is to obtain three important results: (i) a simplified expression of the force-extension relation under isometric condition valid for large values of $N$ and for $\tilde{\beta} \gg 1$, (ii) an explicit expression for the amplitude of the first force peak, representing the nucleation of the unfolded phase, and (iii) a rigorous demonstration of the equivalence of the Gibbs and Helmholtz ensembles for the zipper model. 

Let us then introduce the average chain stretch $\bar \lambda = \frac{x_N}{\ell N}$, prescribed to the chain under isometric condition.
By using Eqs.\eqref{xn_helmholtz_adim} and \eqref{Z_H_compact}, we obtain the force-extension relation in the form
\begin{equation}
\label{lfr}
\langle \tilde{f} \rangle = \frac{
(F_0G_0+F_NG_N)\left(1-e^{-2\tilde{\beta}}\right)+e^{-2\tilde{\beta}}\displaystyle\sum_{\xi=0}^NF(\xi)G(\xi)
}{
(F_0+F_N)\left(1-e^{-2\tilde{\beta}}\right)+e^{-2\tilde{\beta}}\displaystyle\sum_{\xi=0}^NF(\xi)
},
\end{equation}
where we introduced the following functions%
\begin{equation}
\begin{split}
G(\xi)=&\hat{k}\left( \frac{\xi}{N}\right)  \frac{(N-\xi)(\bar \lambda-1)+ \xi(\bar \lambda-\chi)}{N},\\
F(\xi)=&\frac{\exp \left\lbrace -\tilde{\beta}\left[\hat{k}\left( \frac{\xi}{N}\right)\frac{[(N-\xi)(\bar \lambda-1)+ \xi(\bar \lambda-\chi)]^2}{2\,N}+\Delta \tilde{E}\xi\right]\right\rbrace }{\sqrt{N\alpha^{\xi}/\hat{k}\left( \frac{\xi}{N}\right)}},
\end{split}
\end{equation}
we defined the rescaled global stiffness of the system (with $0\leq t\leq 1$)
\begin{equation}
\hat{k}(t)=\left(\frac{1-t}{\tilde{k}}+\frac{t}{\alpha\tilde{k}}\right)^{-1},
\end{equation}
and we used the compact notations $F_0=F(0),G_0=G(0),F_N=F(N),G_N=G(N)$.
 Let us now consider the behavior in the thermodynamical limit $N\rightarrow \infty$.  By following the approach suggested in Refs.\cite{prr,prr_} to obtain explicit analytical results, not only for the stress plateau but also for the stress peak, we may consider the Euler-MacLaurin (EM) approximation for a given function $\phi$
\begin{equation}
\sum_{\xi=0}^N\phi(\xi) \simeq \int_0^N\phi(\xi)\mathrm{d}\xi+\frac{\phi(0)+\phi(N)}{2},
\end{equation}
where higher order terms of the EM approximation would lead to more detailed, but analytically cumbersome results. Thus, from Eq.\eqref{lfr} we get
\begin{equation}\label{langfrang}
\langle \tilde{f} \rangle = \frac{
(F_0G_0+F_NG_N)\left(1-\frac{e^{-2\tilde{\beta}}}{2}\right)+e^{-2\tilde{\beta}}\int_{0}^N F(\xi)G(\xi)\mathrm{d}\xi
}{
(F_0+F_N)\left(1-\frac{e^{-2\tilde{\beta}}}{2}\right)+e^{-2\tilde{\beta}}\int_{0}^N F(\xi)\mathrm{d}\xi
}.
\end{equation}
To simplify the calculation, we may rewrite the integrals as
\begin{align}
\int_0^NF(\xi)\mathrm{d}\xi=&\,\sqrt{N}\int_0^1e^{Ng(\eta)}\sqrt{\hat{k}(\eta)}\mathrm{d}\eta,
\label{intperf}\\
\int_0^NF(\xi)G(\xi)\mathrm{d}\xi=&\,\sqrt{N}\int_0^1e^{Ng(\eta)}\sqrt{\hat{k}(\eta)}f(\eta)\mathrm{d}\eta,\label{intperfg}
\end{align}
where we introduced the phase fraction $\eta=\xi/N$ and the auxiliary functions
\begin{align}
\label{g}
g(\eta)=&%
\frac{\eta}{2}\log\frac{1}{\alpha}-\tilde{\beta}\left(\eta\Delta \tilde{E}+\hat{k}(\eta)\frac{[(1-\eta)(\bar \lambda-1)+ \eta(\bar \lambda-\chi)]^2}{2}\right),\\
\label{f}
f(\eta)=&\hat{k}(\eta)\left[(1-\eta)(\bar \lambda-1)+ \eta(\bar \lambda-\chi)\right].
\end{align}
Now, we use the following general results \cite{ABL,KWB}. Let %
\begin{equation}
 I(x)=\int_a^b e^{\,x\,\,\mathcal{G}(t)}\mathcal{F}(t)\mathrm{d}t,
 \label{intint}
\end{equation}
then, we have

\noindent\textbf{1.} Suppose $\mathcal{F}$ is bounded and continuous on $(a,\,b)$, and $\mathcal{F}(a)\mathcal{F}(b)\neq0$. Suppose also that $\mathcal{G}$ is strictly monotone and differentiable and that $\frac{\mathcal{F}(a)}{\mathcal{G}'(a)}$ and $\frac{\mathcal{F}(b)}{\mathcal{G}'(b)}$ both exist as finite reals, defined as limits if either endpoint is infinite. Assume also that the integral in Eq.~\eqref{intint} exists for all $x>0$. Then, we have that
\begin{equation}
I(x)\underset{x\to\infty}{\sim}\frac{1}{x}\frac{\mathcal{F}(b)}{\mathcal{G}'(b)}e^{\,x\,\,\mathcal{G}(b)}-\frac{1}{x}\frac{\mathcal{F}(a)}{\mathcal{G}'(a)}e^{\,x\,\,\mathcal{G}(a)}.
\label{i}
\end{equation}

\noindent\textbf{2.} Suppose $\mathcal{F}$ is bounded and continuous on $(a,\,b)$, that $\mathcal{G}$ has unique maximum at some $c$ in the open interval $(a,\,b)$, $\mathcal{G}$ is differentiable in some neighborhood of $c$, $\mathcal{G}''(c)$ exists and is $\mathcal{G}''(c)<0$, and that $\mathcal{F}(c)\neq 0$. Then, we have that
\begin{equation}
I(x)\underset{x\to\infty}{\sim}\frac{\sqrt{2\pi}\mathcal{F}(c)e^{\,x\,\,\mathcal{G}(c)}}{\sqrt{-x\mathcal{G}''(c)}}.
\label{ii}
\end{equation}

We use the results in Eqs.\eqref{i} and \eqref{ii} to analyze the behavior of the integrals defined in Eqs.\eqref{intperf} and \eqref{intperfg}.  First, by using Eq.\eqref{g}, we obtain
\begin{equation}
\begin{split}
\frac{\partial g}{\partial\eta}=&\tilde{\beta}\hat{k}(\eta)\left[(1-\eta)(\bar \lambda-1)+ \eta(\bar \lambda-\chi)\right](\chi-1)-\tilde{\beta}\Delta \tilde{E}+\frac{1}{2}\log\frac{1}{\alpha}\\
&+\frac{\tilde{\beta}[\hat{k}(\eta)]^2}{2}[(1-\eta)(\bar \lambda-1)+ \eta(\bar \lambda-\chi)]^2\left(\frac{1}{\alpha\tilde{k}}-\frac{1}{\tilde{k}}\right).
\end{split}
\end{equation}
Solving $\frac{\partial g}{\partial\eta}=0$, we obtain the solution $\eta_0$ that represents the stationary point to be computed. Thus, following the definition in Eq.(\ref{f}), we also introduce the value
\begin{equation}
\label{f(eta)}
\tilde{f}_0=\tilde{f}(\eta_0)=\hat{k}(\eta_0)\left[(1-\eta_0)(\bar \lambda-1)+ \eta_0(\bar \lambda-\chi)\right].
\end{equation}
We observe that the equation $\frac{\partial g}{\partial\eta}=0$ can be written in terms of $\tilde{f}_0$ as
\begin{equation}
\label{MF_stationary}
\frac{1}{2}\log\left(\frac{1}{\alpha}\right)-\tilde{\beta}\left[\Delta\tilde{E}-(\chi-1)\tilde{f}_0-\left(\frac{1}{\alpha}-1\right)\frac{\tilde{f}_0^2}{2\tilde{k}}\right]=0,
\end{equation}
that coincides with Eq.\eqref{MF} (\textit{i.e.} with $\delta=0$). Thus, we identify $\tilde{f}_0$ with the Maxwell force $\tilde{f}_M$. Now, we can use Eq.\eqref{ii} (stationary phase theorem) to simplify Eqs.\eqref{intperf} and \eqref{intperfg} only if $0<\eta_0<1$ (the stationary point must be within the integration interval). For $\eta_0=0$, we have $\tilde{f}_0=(\bar \lambda-1)\tilde{k}$, and for $\eta_0=1$, we have $\tilde{f}_0=(\bar \lambda-\chi)\alpha\tilde{k}$, as given by Eq.\eqref{f(eta)}. Hence, the interval $0<\eta_0<1$ is equivalent to $1+\frac{\tilde{f}_0}{\tilde{k}}<\bar \lambda<\chi+\frac{\tilde{f}_0}{\alpha \tilde{k}}$, which corresponds to the force plateau region between the two elastic branches. Therefore, only in the plateau interval we have a stationary point and we can approximate Eqs.\eqref{intperf} and \eqref{intperfg} with Eq.\eqref{ii}. The application of the stationary phase method is further justified by the relation $g''(\eta_0)<0$, simply proved by a direct evaluation
\begin{equation}
g''(\eta_0)=
-\tilde{\beta}\hat{k}(\eta_0)\left[\left( \chi+\frac{\tilde{f}_0}{\alpha \tilde{k}}\right) -\left(1+\frac{\tilde{f}_0}{\tilde{k}}\right)\right]^2<0.
\end{equation}
Note that from now on, we use the notation $\tilde{f}_M$ for the quantity $\tilde{f}_0$, to be consistent with previous Sections. We obtain from Eq.(\ref{langfrang}) the expression
\begin{equation}
\langle \tilde{f} \rangle=
\frac
{\mathcal{C}_N\left[e^{\,\Delta g_0 N}(\bar \lambda-1)\tilde{k}^{\frac{1}{2}}+e^{\,\Delta g_1 N}(\bar \lambda-\chi)(\alpha\tilde{k})^{\frac{3}{2}}\right]+\tilde{f}_M}
{\mathcal{C}_N\left[\sqrt{\tilde{k}}\,\,e^{\,\Delta g_0 N}+\sqrt{\alpha\tilde{k}}\,\,e^{\,\Delta g_1 N}\right]+1},
\label{staph}
\end{equation}
where
\begin{align}
\Delta g_0=&g(0)-g(\eta_0)=-\frac{\tilde{\beta}\tilde{k}}{2}\left[\bar \lambda-\left(1+\frac{\tilde{f}_M}{\tilde{k}}\right)\right]^2,\\
\Delta g_1=&g(1)-g(\eta_0)=-\frac{\tilde{\beta}\alpha\tilde{k}}{2}\left[\bar \lambda-\left(\chi+\frac{\tilde{f}_M}{\alpha\tilde{k}}\right)\right]^2,\\
\mathcal{C}_N=&\frac{2 \,e^{\,2\tilde{\beta}}-1}{2 }\sqrt{\frac{\tilde{\beta}}{2\pi\,N}}\left[\chi+\frac{\tilde{f}_M}{\alpha \tilde{k}}-\left(1+\frac{\tilde{f}_M}{\tilde{k}}\right)\right].
\label{ccc}
\end{align}
This important analytical expression for the force-extension relation contains all the physical features describing the Helmholtz ensemble, as the first nucleation peak and the coalescence one. In particular, it can be used to evaluate these peaks as function of $N$ and the temperature $T$. Moreover, considering that $\Delta g(0)<0$ and $\Delta g(1)<0$ (since $g(\eta_0)$ is the maximum value of $g(\eta)$ in the interval $0<\eta<1$), we may prove that in the thermodynamic limit
\begin{equation}
\lim_{N\to\infty}\langle \tilde{f}\rangle = \tilde{f}_M, \label{appr}
\end{equation} 
which is valid for $1+\frac{\tilde{f}_M}{\tilde{k}}<\bar \lambda<\chi+\frac{\tilde{f}_M}{\alpha\tilde{k}}$, and meaning that the Maxwell force is the same for both the Helmholtz and Gibbs ensembles, proving their equivalence for the zipper model. 

%----------------------------------------------------
%
\begin{figure}[t]
    \centering
    \includegraphics[scale=0.9]{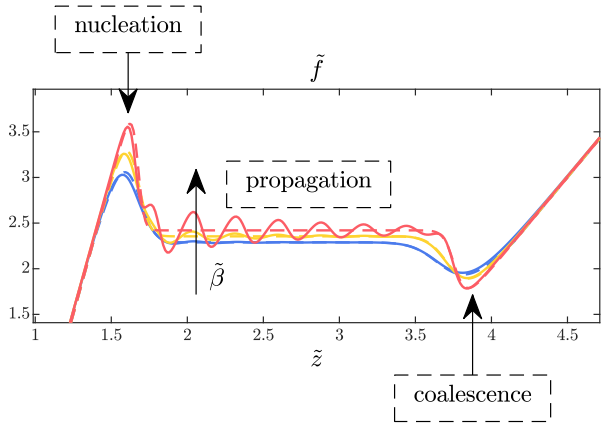}
    \caption{Comparison between the isometric force-extension response of the exact zipper model, obtained by the partition function in Eq.\eqref{Z_H_compact} (solid lines), and the result of the stationary phase method, stated in Eq.\eqref{staph} (dashed lines). We considered different values of the Ising dimensionless parameter $\tilde{\beta}=\{1, 1.5, 3\}$. Adopted parameters: $N=10$, $\tilde{k}=6$, $\tilde{h}=2$, $\alpha=1/3$, $\tilde{Q}(-1)=0$, $\tilde{Q}(+1)=\Delta \tilde{E}=6$, $\chi=3$.}
    \label{fig:isingphst_plot}
\end{figure}
%
%----------------------------------------------------

In order to conclude the analysis, we have to simplify Eq.\eqref{langfrang} also for the external regions, \textit{i.e.} for $\bar \lambda<1+\frac{\tilde{f}_M}{\tilde{k}}$ and $\bar \lambda>\chi+\frac{\tilde{f}_M}{\alpha\tilde{k}}$.
Hence, we suppose that the critical point $\eta_0$ is external to the interval $(0,1)$. We can have either $\eta_0<0$ or $\eta_0>1$.
In these cases, the asymptotic behavior of the integrals in Eqs.\eqref{intperf} and \eqref{intperfg} is described by Eq.\eqref{i}.
Then, Eq.\eqref{langfrang} assumes the form
\begin{equation}
\langle \tilde{f}\rangle=%
\frac%
{e^{Ng(0)}\tilde{k}^{\frac{3}{2}}(\bar \lambda-1)+e^{Ng(1)}(\alpha\tilde{k})^{\frac{3}{2}}(\bar \lambda-\chi)+\frac{2}{2e^{2\tilde{\beta}}-1}\mathcal{A}}%
{e^{Ng(0)}\sqrt{\tilde{k}}+e^{Ng(1)}\sqrt{\alpha\tilde{k}}+\frac{2}{2e^{2\tilde{\beta}}-1}\mathcal{B}},
\label{lfrcasei}
\end{equation}
where
\begin{align}
\mathcal{A}=&\frac{(\alpha\tilde{k})^{\frac{3}{2}}}{g'(1)}e^{Ng(1)}(\bar \lambda-\chi)-\frac{\tilde{k}^{\frac{3}{2}}}{g'(0)}e^{Ng(0)}(\bar \lambda-1),\\
\mathcal{B}=&\frac{\sqrt{\alpha\tilde{k}}}{g'(1)}e^{Ng(1)}-\frac{\sqrt{\tilde{k}}}{g'(0)}e^{Ng(0)}.
\end{align}
Now, by using Eq.\eqref{MF_stationary}, we determine the quantity $g(1)-g(0)$ that eventually reads
\begin{equation}
g(1)-g(0)=\frac{\tilde{\beta}\tilde{k}}{2}\left[\left(\bar \lambda-1-\frac{\tilde{f}_M}{\tilde{k}}\right)^2 -\alpha\left(\bar \lambda-\chi-\frac{\tilde{f}_M}{\alpha\tilde{k}}\right)^2\right].
\end{equation}
Accordingly, in the limit of $N\to\infty$ we obtain 
\begin{equation}
    g(1)-g(0)
    \left\{
    \begin{aligned}
        &>0 		\,\,\,\,\,\,	\text{if}	\,\,\,\,\,\,	 \bar \lambda>\chi+\frac{\tilde{f}_M}{\alpha\tilde{k}}\Rightarrow \langle \tilde{f}\rangle =\alpha\tilde{k} (\bar \lambda-\chi),\\
        &<0		\,\,\,\,\,\,	\text{if}	\,\,\,\,\,\,	 \bar \lambda<1+\frac{\tilde{f}_M}{\tilde{k}}\Rightarrow\langle \tilde{f}\rangle =\tilde{k} (\bar \lambda-1),\\
\end{aligned}
    \right.
    \label{elasticbranches}
\end{equation}
 a result representing  the elastic branches in the external regions, corresponding to the fully folded (left) and fully unfolded (right) phases. This completes the proof of the equivalence of the ensembles in the thermodynamic limit for the zipper model.

\begin{figure}[t]
    \centering
    \includegraphics[scale=0.9]{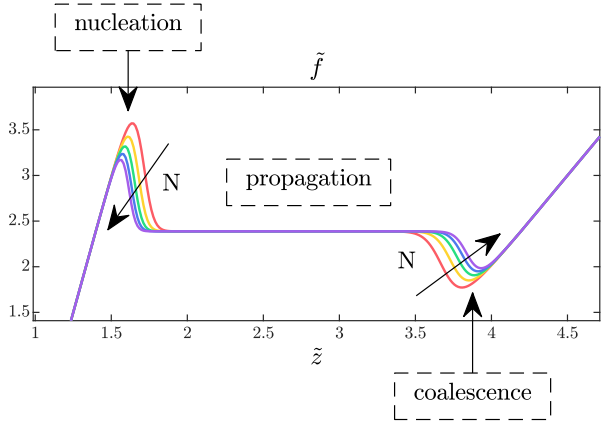}
    \caption{Evolution of the isometric force-extension curve, stated in Eq.\eqref{staph} as obtained by the stationary phase method, with an increasing number $N$ of units. Adopted parameters: $\tilde{k}=6$, $\tilde{h}=2$, $\alpha=1/3$, $\tilde{Q}(-1)=0$, $\tilde{Q}(+1)=\Delta \tilde{E}=6$, $\chi=3$, $\tilde{\beta}=2$, $N=\{8, 10, 12, 14, 16\}$.}
    \label{fig:Hel2Gib}
\end{figure}
%
%----------------------------------------------------

In Fig.\ref{fig:isingphst_plot}, we compare the force-extension curves obtained through the exact zipper model with the ones delivered by the stationary phase method. Observe that while this approximation, stated in Eqs.\eqref{staph} and \eqref{lfrcasei}, is finer as $N$ grows, our results show the possibility of a good approximation of just the nucleation peak already for small values of $N$ (we used $N=10$ in Fig.\ref{fig:isingphst_plot}). Also, the approximation correctly reproduces the last downwards peak, corresponding to the coalescence of the remaining folded regions to the fully unfolded phase. 
The most important result, quite evident in this framework, is the dependence of the Maxwell (propagation) force on the temperature, through the parameter $\tilde{\beta}$.
We observe that for the represented case with $\alpha <1$ (softening) as $\tilde{\beta}$ decreases ({\it e.g.} temperature increases), the value of the propagation force $f_M$ decreases, also affecting the height and the width of the nucleation and coalescence peaks with an entropic stabilization of the unfolded phase.

In Fig.\ref{fig:Hel2Gib}, we observe that, when the number of units $N$ increases and the thermodynamic limit is approached, the mechanical response of the system in the Helmholtz ensemble (see Eq.\eqref{staph}) approaches the Gibbs behavior, where there is a force plateau without nucleation and coalescence force peaks.
In particular, we notice that by means of the stationary phase method (when $N\to \infty$) the typical saw-tooth path of the isometric condition disappears whereas only the first and the last peaks remain, as expected from a macroscopic point of view.
We remark that, in experiments and simulations of phase transformations in nanowires or other nanostructures, it is possible to observe only the first nucleation peak, while the last one is rarely attained because of the breaking of the specimen or because of the boundary effects induced by the grips of the traction device.

\begin{figure}[t]
    \centering
    \includegraphics[scale=0.9]{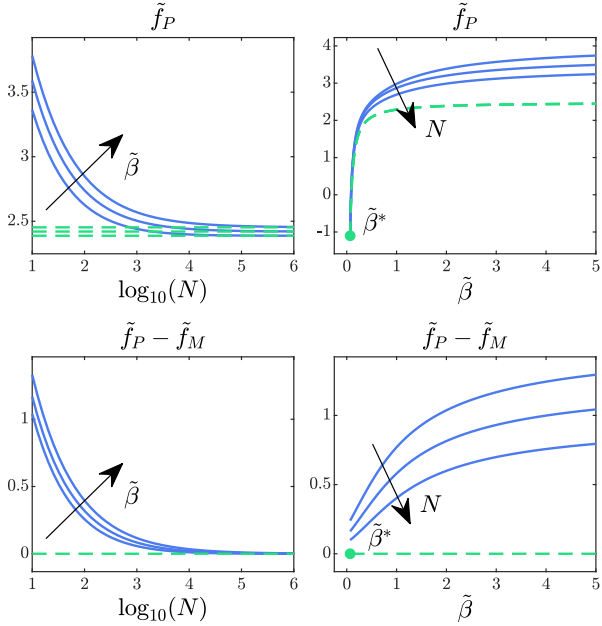}
    \caption{Behavior of the nucleation force peak as function of the number $N$ of units of the chain and of $\tilde{\beta}$. First row: $\tilde{f}_P$ versus $\log_{10}(N)$, parametrized by $\tilde{\beta}=\{2, 3, 6\}$, and $\tilde{f}_P$ versus $\tilde{\beta}$, parametrized by $N=\{10, 15, 25\}$. The green dashed lines correspond to $\tilde{f}_M$, and $\tilde{\beta}^*$ is the critical value corresponding to $\tilde f_F$ (floor force with $\alpha<1$). Second row:  $\tilde{f}_P-\tilde{f}_M$ versus $\log_{10}(N)$, parametrized by $\tilde{\beta}=\{2, 3, 6\}$, and $\tilde{f}_P-\tilde{f}_M$ versus $\tilde{\beta}$, parametrized by $N=\{10, 15, 25\}$. Here, the green dashed lines correspond to zero. We adopted the parameters $\tilde{k}=6$, $\tilde{h}=2$, $\tilde{Q}(-1)=0$, $\tilde{Q}(+1)=\Delta \tilde{E}=6$, and $\chi=3$.}
    \label{fig:fMonN}  
\end{figure}

\begin{figure}[t]
    \centering
    \includegraphics[scale=0.9]{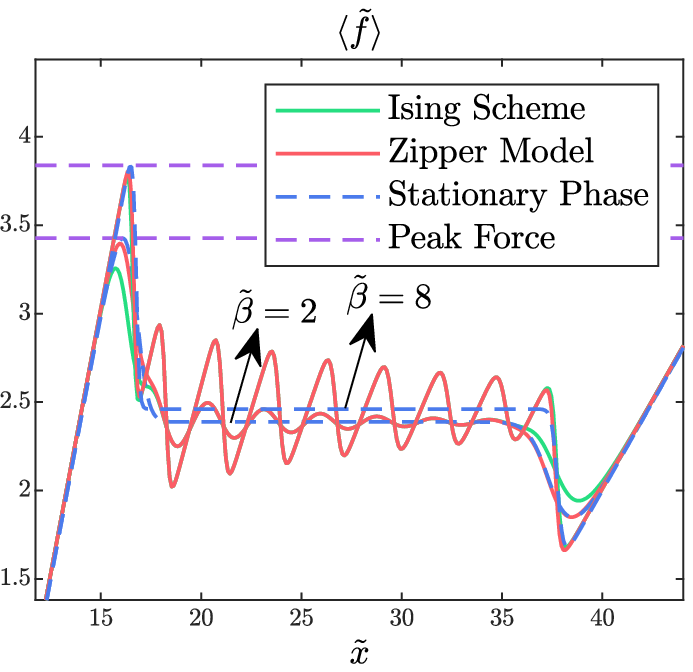}
    \caption{Comparison of the force-extension relation observed within the Helmholtz ensemble and obtained with (i) the Ising scheme, (ii) the zipper model, and (iii) the stationary phase approximation. Also the amplitude of the first nucleation peak calculated by Eq.\eqref{eq:fp-fm} is shown. Adopted parameters: $N=10$, $\tilde{k}=6$, $\tilde{h}=2$, $\tilde{Q}(-1)=0$, $\tilde{Q}(+1)=\Delta \tilde{E}=30$, $\chi=3$, $\tilde{\beta}=\{2, 8\}$.}
    \label{fig:esatto_zipper_fasestaz}
\end{figure}

Given the importance of the first nucleation force peak, we now look for an explicit expression of its magnitude. Consider Eq.\eqref{staph} and observe that in the region of the first peak only the term $e^{\,\Delta g_0 N}$ is relevant, being $e^{\,\Delta g_1 N}$ negligible. Then, we can write a reduced form of Eq.\eqref{staph}, as follows
\begin{equation}
\langle \tilde{f}\rangle = \tilde{k}\frac{\sqrt{\tilde{k}}\,\mathcal{C}_N\,e^{-\frac{\tilde{\beta}\tilde{k}}{2} N\mathrm{y}^2}\left(\mathrm{y}+\frac{\tilde{f}_M}{\tilde{k}}\right)+\tilde{f}_M}{\sqrt{\tilde{k}}\,\mathcal{C}_N\,e^{-\frac{\tilde{\beta}\tilde{k}}{2} N\mathrm{y}^2}+1},
\end{equation}
where we introduced  the change of variable $\mathrm{y}=\bar \lambda-1-\frac{\tilde{f}_M}{\tilde{k}}$. Now, let us consider as reference for the force its Maxwell value, so as to obtain
\begin{equation}
\langle \tilde{f}\rangle - \tilde{f}_M = \tilde{k}\mathrm{y}\frac{\sqrt{\tilde{k}}\,\mathcal{C}_N\,e^{-\frac{\tilde{\beta}\tilde{k}}{2} N\mathrm{y}^2}}{\sqrt{\tilde{k}}\,\mathcal{C}_Ne^{-\frac{\tilde{\beta}\tilde{k}}{2} N\mathrm{y}^2}+1}.
\label{jump}
\end{equation}
We search for the stationary point of Eq.~\eqref{jump} by derivation, easily obtaining 
\begin{equation}
\left(\frac{\tilde{\beta}\tilde{k}}{2} N\mathrm{y}^2-\frac{1}{2}\right)\,e^{\,\frac{\tilde{\beta}\tilde{k}}{2} N\mathrm{y}^2-\frac{1}{2}}=\frac{\mathcal{C}_N}{2}\sqrt{\frac{\tilde{k}}{e}},
\end{equation}
which is a Lambert equation of the form $\mathrm{w}e^\mathrm{w}=\mathrm{s}$, solved by $\mathrm{w}=\mathrm{W}_0(\mathrm{s})$, where $\mathrm{W}_0$ is the Lambert function (more precisely, $\mathrm{W}_0$ is the principal branch of the Lambert function) \cite{lambert,www}. 
Thus, we may write 
\begin{equation}
\frac{\tilde{\beta}\tilde{k}}{2} N\mathrm{y}^2-\frac{1}{2}= \mathrm{W}_0\left(\frac{\mathcal{C}_N}{2}\sqrt{\frac{\tilde{k}}{e}}\right).
\label{lambsol}
\end{equation}
Eventually, by using Eq.~\eqref{lambsol} in Eq.\eqref{jump} we get
\begin{equation}
\label{eq:fp-fm}
\tilde{f}_P-\tilde{f}_M=\frac{\tilde{k}}{\sqrt{\frac{\tilde{\beta}\tilde{k}}{2} N}}\frac{\mathrm{W}_0\left(\frac{\mathcal{C}_N}{2}\sqrt{\frac{\tilde{k}}{e}}\right)}{\sqrt{\frac{1}{2}+\mathrm{W}_0\left(\frac{\mathcal{C}_N}{2}\sqrt{\frac{\tilde{k}}{e}}\right)}},
\end{equation}
where $\tilde{f}_P$ is the value of the nucleation force peak. It is worth to remark that Eq.\eqref{eq:fp-fm} gives the explicit dependence of the force peak on both $\tilde{\beta}$ and the number of units of the chain $N$.

In the first row of Fig.\ref{fig:fMonN} we study the nucleation peak with respect to $N$ (parametrized by $\tilde{\beta}$) and with respect to $\tilde{\beta}$ (parametrized by $N$).
Similarly, in the second row, we show the plots of $\tilde{f}_P-\tilde{f}_M$. The trend can be summarized by stating that $\tilde{f}_P-\tilde{f}_M$ increases for increasing values of $\tilde{\beta}$ and decreases for increasing values of $N$. Moreover, $\tilde{f}_P$ converges to $\tilde{f}_M$ in the thermodynamic limit. 

The consistency of the approaches presented in this work can be appreciated by observing Fig.\ref{fig:esatto_zipper_fasestaz}.
Here, we compare the force-extension relation evaluated within the Helmholtz ensemble and obtained with (i) the complete Ising scheme discussed in Section \ref{Helmholtz}, (ii) the simplified zipper model introduced in Section \ref{Zipper-Helmholtz}, and (iii) the stationary phase approximation given in Eq.\eqref{staph}.
Moreover, we also show the amplitude of the first peak calculated by Eq.\eqref{eq:fp-fm}.
This plot proves that, for a sufficiently large value of $N$ and with the hypothesis of strong ferromagnetic interactions (large values of $\tilde{\beta}$), the three frameworks give the same results and hence the hypothesis introduced are consistent with the problem under investigation.

%---------------------------------------------------------------------------------------------------------------------------------------
\section{Applications to the tensile behavior of nanowires}

While, as pointed out in the Introduction, our model can be applied to different multistable systems with softening or hardening effects ({\it i.e.} with different stiffness of the different phases) and non local interactions, we here focus our attention on the specific application to phase transition in metallic nanowires. We can show the ability of the model to predict the main physical phenomena observed during the phase transition at variable temperature.  
Single-crystalline metal nanowires with nanometric cross-section can exhibit a pseudo-elastic behavior characterized by very large eleongations which can be up to 50\% of the original length \cite{dmo0b,dmo0c,dmo1,dmo2}. This behavior, which is exceptional compared to all other shape memory alloys, is due to a reversible lattice reorientation process with twin boundary propagation between two differently oriented face-centered cubic (FCC) crystalline structures. This behavior is typical of Copper  (Cu) and Nickel (Ni), where we can identify the two following configurations: the original one named $<110>/\left\lbrace 111\right\rbrace $  with axis $<110>$  and surfaces $\left\lbrace 111\right\rbrace$, and the deformed one named $<001>/\left\lbrace 100\right\rbrace $ with axis $<001>$ and surfaces $\left\lbrace 100\right\rbrace$, as shown \textit{e.g.} in Figs.2 and 3 of Ref.\cite{dmo0c}. 

To compare the behavior with our one dimensional system, let $s$ be the area on the cross-section pertaining to a single longitudinal chain of atoms in the nanowire crystal structure. Let then $M$ the average number of atoms in the cross section so that the total area is given by $S=Ms$. Similarly, we use $\ell$ to indicate roughly the lattice constant of the crystal structure and hence $L=N\ell$ is the total nanowire length, being $N$ the number of atoms in the longitudinal direction. Accordingly, $SL$ is the total volume of the system, $MN$ is the total number of atoms, $s\ell$ is the average volume pertaining to one atom, and one can observe also that $s\simeq\ell^2$ and $s\ell\simeq\ell^3$.

It is important to remark that both the Maxwell force in Eqs.\eqref{MF} and the force-extension relation in Eq.\eqref{staph}, have been obtained for a single chain of units (here atoms) and therefore the continuous parameters must be introduced as follows. We define
\begin{equation}
E_f=\frac{k_0\ell}{s} \,\,\,\text{[GPa]}, \qquad E_u=\frac{h_0\chi\ell}{s}\,\,\,\text{[GPa]},
\end{equation}
the Young moduli of the folded ($f$) and unfolded ($u$) phases, respectively. Accordingly, $k_0$ and $h_0$ are the elastic constants of the two crystals in the harmonic limit. Moreover, the jump energy between the two configurations can be written as $\Delta E= s\ell\Delta e$ where $\Delta e$ $[$J/m$^3]$ is the energy density difference between the two crystal states. In addition, the Ising energy can be rewritten as $J=\Lambda\, s$, where $\Lambda$ $[$J/m$^2]$ is the surface energy density of the twin boundary separating original and deformed crystals. We remark that the parameters $\Delta e$ and $\Lambda$ may depend on the total area $S$ when elastic surface effects are relevant, as in the case of metallic nanowires. Further, the longitudinal (normal) stress is defined by $\sigma=\langle f\rangle/s$, with $\langle f\rangle$ given in Eq.\eqref{staph} if we consider the stationary phase approximation. The strain is defined as $\bar \varepsilon=\frac{x-N\ell}{N\ell}$, and therefore we have $\bar \varepsilon=\bar \lambda-1$. 

\begin{figure}[t]
    \centering
    \includegraphics[scale=0.9]{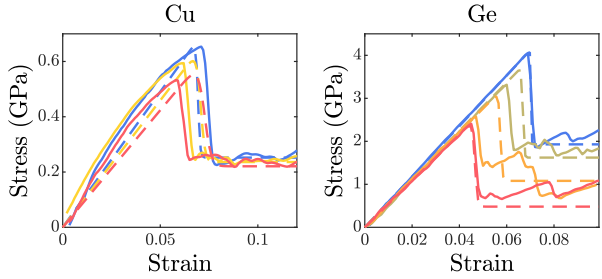}
    \caption{Comparison of the stress-strain curves obtained with molecular dynamic simulations (solid lines) and through our theoretical approach (dashed lines) for Cu nanowires (1.96 nm $\times$ 1.96 nm) in left panel and Ge nanowires (radius of 1.8 \AA) in right panel. In left panel, we used the temperatures  $T$=100 K (blue lines), $T$=200 K (yellow lines), and $T$=300 K (red lines). In right panel, we used the temperatures $T$=100 K (blue lines), $T$=200 K (green lines), $T$=300 K (yellow lines), and $T$=400 K (red lines).}
    \label{md}
\end{figure}

We use the introduced continuous quantities to rewrite previous expressions, for the purpose of mimicking the tensile behavior of one-dimensional structures. Indeed, Eqs.\eqref{MF} and \eqref{eq:delta} can be rewritten as
\begin{eqnarray}
\frac{1}{2}\log\left( \frac{E_f}{E_u}\chi\right) +\beta s\ell\Biggl[ \sigma_M(\chi-1)+\frac{1}{2}\sigma_M^2\left(\frac{\chi}{E_u}-\frac{1}{E_f}\right)-\Delta e\Biggr] =0,
\end{eqnarray}
that is used to obtain the temperature dependent Maxwell stress $\sigma_M$. We observe that, due to the hypothesis of homogeneity in the cross-section, the wire specific response coincides with that of a longitudinal chains of atoms. Correspondingly, the nucleation stress can be defined, using Eq.\eqref{eq:fp-fm} as follows
 \begin{equation}
\label{eq:sp-sm}
\sigma_P-\sigma_M=\sqrt{\frac{2 E_f}{\beta s\ell N}}\frac{\mathrm{W}_0\left(C\right)}{\sqrt{\frac{1}{2}+\mathrm{W}_0\left(C\right)}},
\end{equation}
where $C=\frac{\mathcal{C}_N}{2}\sqrt{\frac{\tilde{k}}{e}}$ is, by using \eqref{ccc}, given by
 \begin{equation}
C=\frac{2\,e^{2\beta\Lambda s}-1}{2}\left[ \chi-1+\sigma_M\left(\frac{\chi}{E_u}-\frac{1}{E_f}\right)\right] \sqrt{\frac{\beta s\ell E_f}{2 eN}}.
\end{equation}
Moreover, by recalling the definition of $\beta=1/(K_B T)$, we introduce also the the parameterizations
\begin{equation}
\label{paras}
\Delta e=\gamma\frac{K_B T_0}{s\ell}, \qquad\Lambda=\tilde \beta_o \frac{K_B T_0}{s}
\end{equation}
to obtain the two values of the characteristic energies. Here, we use the reference temperature $T_0=300$ K.

In Ref.\cite{dmo0c} a series of molecular dynamics simulations have been performed to obtain the stress-strain behavior of Cu nanowires at different temperatures and under quasi-static tensile deformation, which is compatible with our equilibrium statistical mechanics approach. The simulations are based on the embedded-atom-method interatomic potential for Cu~\cite{dmo0c}, and the uniaxial displacement-controlled loading strategy is applied coherently with our Helmholtz ensemble. The stress-strain behaviors of this Cu nanowire (1.96 nm $\times$ 1.96 nm section) at temperatures of $100, 200$, and $300$ K is compared with our model in Fig.\ref{md}, left panel. The theoretical results are based on the stationary phase approximation stated in Eq.\eqref{staph}. As one can see, the model is able to reproduce the magnitude of both the nucleation stress and the Maxwell stress plateau. In Table~\ref{tab1}, one can find the temperature dependent Young moduli we adopted to reproduce the simulations of Ref.\cite{dmo0c}. Moreover, we adopt the following parameters: $N=206$, $\ell=0.14$ nm, $s=\ell^{2}$ and $\chi=4.4$.
By introducing the parameterizations in Eq.(\ref{paras}), we obtained the adimensional values $\gamma=7.7$ and $\tilde \beta_0=6.8$. We remark that since $\tilde \beta_0$, representing our non dimensional parameter $\tilde \beta$ for the higher value of the temperature, is sufficiently larger than 1, the use of the zipper model, simplified by the stationary phase method, is justified. Moreover, the propagation of a single twin boundary is confirmed by the simulations in Ref.\cite{dmo0c}.

%------------------------------------------------
%
\begin{table}
\centering
\begin{tabular}{cc}
\toprule
Young modulus & Value \\
\midrule
$E_f(T=100 K)$      	& $122$   GPa  	\\
$E_f(T=200 K)$      	& $114$   GPa  	\\
$E_f(T=300 K)$      	& $105$   GPa  	\\
$E_u(T=100 K)$     	& $10$   GPa  	\\
$E_u(T=200 K)$     	& $11$   GPa  	\\
$E_u(T=300 K)$     	& $12$   GPa  	\\
\bottomrule
\end{tabular}
\caption{Mechanical parameters used to reproduce the molecular dynamics simulations on a Cu nanowire of Ref.\cite{dmo0c}.\label{tab1}}
\end{table}
%
%------------------------------------------------

We propose also a second example of application of our model to ultrathin semiconductor Germanium (Ge) nanowires that have been recently studied due to their peculiar properties. Here, we compare our results with molecular dynamics simulations of the helix Ge nanowire with radius of 1.8 \AA (see Fig.1 in Ref.\cite{dmo5}). The stress-strain curves has been obtained for temperatures of $100, 200, 300$, and $400$ K~\cite{dmo5}. The simulations for germanium were based on the Stillinger-Weber potential, coupled with the Nos\'{e}-Hoover thermostat to impose the system temperature.
In Fig.~\ref{md}, right panel, we show the comparison of the simulations with our model, and we observe that the nucleation and Maxwell stresses are in fairly good agreement. Unfortunately, the numerical stress plateaus are rather noisy because they correspond to a single simulation (without averaging) and the amorphous phase is strongly fluctuating. Anyway, the temperature dependent behavior of such plateaus is clearly visible. The helical structure is perfectly ordered initially (see Fig.1 in Ref.\cite{dmo5}), and shows a transition to an amorphous structure (see Fig.5 in Ref.\cite{dmo5}) following a transition process that begins with a stress peak, namely the nucleation stress. After this nucleation, the unfolded region evolves `smoothly' into the one-atom chain structure. In our theoretical formulation, we used the parameters listed in Table~\ref{tab2} corresponding to the data in Ref.~\cite{dmo5}. Moreover, we adopted the parameters: $N=5562$, $\ell=0.208$ nm, $s=\ell^{2}$ and $\chi=3.36$. In particular, by introducing the parameterizations in Eq.(\ref{paras}), we obtained the adimensional values $\gamma=2.3$ and $\tilde\beta_0=25.3$ (or $\tilde \beta =19$ in the case of highest temperature $T=400 K$) so that the assumption of the zipper model is widely justified. 

%------------------------------------------------
%
\begin{table}
\centering
\begin{tabular}{cccc}
\toprule
Young modulus & Value  \\
\midrule
$E_f(T=100 K)$      	& $588$   GPa  	\\
$E_f(T=200 K)$      	& $561$   GPa  	\\
$E_f(T=300 K)$      	& $552$   GPa  	\\
$E_f(T=400 K)$      	& $540$   GPa  	\\
$E_u(T=100 K)$     	& $23$   GPa  	\\
$E_u(T=200 K)$     	& $43$   GPa  	\\
$E_u(T=300 K)$     	& $63$   GPa  	\\
$E_u(T=400 K)$     	& $83$   GPa  	\\
\bottomrule
\end{tabular}
\caption{Mechanical parameters used to reproduce the molecular dynamics simulations on a Ge ultra thin nanowire of Ref.\cite{dmo5}.}
\label{tab2}
\end{table}
%
%------------------------------------------------

To sum up our results, we considered here systems where two possible configurations exist. In the first application they correspond to two different crystal structures whereas in the second example to a regular helical structure and an irregular amorphous structure. It is worth noticing that the model based on statistical mechanics here developed is able to reproduce the complex behavior of these systems. 
It is important to highlight that in our model there is no crystallographic or morphological information of the three-dimensional structures, whereas the informations of the microstructure are embedded in the  elastic and energetic properties of the system. Thus the model captures the main physical behavior in term of nucleation and propagation stress and their dependence on temperature, without exploiting the microstructural details. 
Nevertheless, the generality of the model allows its application to different systems ranging from material science to cooperative biological structures. Of course, if it is necessary to take into account crystallographic details, structural anisotropies, twin-boundary geometries and other morphological features, then it is necessary to turn to another class of models that are specifically adapted to the problem at hand, \textit{e.g.} the model in Ref.\cite{dmo0c} for the pseudo-elasticity of metallic nanowires. 
On the other hand, the possibility of deducing explicit equations of the temperature dependence of the nucleation and propagation stress, depending on parameters with a clear physical interpretation, represents in our opinion the main advantage of the proposed model both in the perspectives of interpreting the transition behavior of biological materials and phase transition alloys and in the field of new material design.

%---------------------------------------------------------------------------------------------------------------------------------------
\section{Conclusions}

In this work, we elaborated some models, with different level of complexity, to describe the temperature-dependent behavior of  one-dimensional non-local non-convex systems. 
The paradigmatic system under investigation is composed of a sequence of units, exhibiting a bistable behavior described by a two-state potential energy, and which are in interaction with each other. 
On the one hand, this scheme is able to represent the specific feature of several macromolecules of biological origin (mainly proteins). 
In this case, the non-convexity (bistability) describes the possible switching of each domain of the macromolecule between its folded and unfolded states.  In addition, the non-locality, introduced by means of the Ising scheme, depicts the cooperativity observed in most of the biological structures.
From the other hand, the proposed model is equally able to represent the phase transformations in materials. The two states of the bistable behavior represent in this case  two microstructures of the solid material, corresponding \textit{e.g.} to two crystal structures or to austenitic and martensitic phases. This enables the study of pseudo-elasticity and shape memory effects in solid systems such as whiskers, nanowires or nanocomposites. For these systems, the Ising interaction scheme reproduces the energy cost of creating an interface between the two different states of matter
due to the absence of kinematic compatibility.   
Our model then describes in a prototypical way the origin of the nucleation, propagation and, possibly, coalescence phenomena and their possible dependence on temperature when low dimensions are considered as in nanowires so that entropic contributions can compete with bulk elastic energy terms. 

Indeed, in the proposed approach, the thermal effects are carefully considered by introducing the statistical mechanics analysis of the problem, based on the method of the spin variables. This allows us to develop a complete thermodynamic framework of the transformation processes in systems with non-locality and non-convexity. 
The first proposed approach is the most refined and implements the complexity of the system in its entire generality, without simplifying assumptions. The system is constituted by a chain of bistable elements interacting through an Ising scheme. We can study both the Gibbs (isotensional) and the Helmholtz (isometric) ensembles of the statistical mechanics by evaluating the pertinent partition functions in closed form. These results are valid for both the ferromagnetic (positive Ising interaction) and antiferromagnetic cases (negative Ising interaction). However, in this work, we mainly focus on the ferromagnetic behavior.
In this model, each unit can freely assume one of the two admitted states, depending on the system temperature and on the mechanical actions applied to the system. The number of interfaces between folded and unfolded regions is therefore free to vary and represents a measure of the entropic effects on the system. Hence, we can observe an increase in the number of interfaces with increasing temperature for a ferromagnetic system.
The important point is that for strongly ferromagnetic systems (with high energetic cost of interface generation), which are the most common in practical applications, only one interface is observed between the two different states and it propagates along the chain when the traction or elongation are applied to the system. 
In addition to the interface propagation phenomenon, in the isometric Helmholtz case, we can also observe an initial upward peak of force, representing the nucleation of the new unfolded phase, and a final downward peak of force, representing the coalescence of the folded phase into the unfolded one. An important feature of this system is that the force plateau, describing the interface propagation within the chain is in general temperature dependent for both isotensional and isometric boundary conditions. The origin of this dependence lies in the difference between the elastic constants of the unfolded and folded phases. In fact, when these elastic constants are equal, the force plateau becomes independent of temperature and known results from the literature are retrieved.  

Since the thermodynamic limit analysis is extremely complex for the exact Ising model (in particular, for the isometric case), and since the important case for applications is the strongly ferromagnetic one, we have introduced the zipper model where we consider only one interface propagating along the system.  
This assumption makes the calculation of the partition function much simpler in both the Gibbs and the Helmholtz ensembles. The thermodynamic limit for the Gibbs ensemble can be analyzed straightforwardly and the results are perfectly coherent with the exact Ising model studied previously.  
In particular, the temperature dependent behavior of the force plateau is confirmed also by means of this zipper approach.

The investigation of the thermodynamic limit concerning the zipper model under isometric condition, \textit{i.e.} within the Helmholtz ensemble,  represents the third approach of this work. This analysis is based on the stationary phase method and it is useful to obtain three important results: (i) a simplified expression of the force-extension relation under isometric condition  valid for very long and strongly ferromagnetic chains (which perfectly describes the nucleation, propagation and coalescence phenomena), (ii) an explicit expression for the amplitude of the first force peak (based on the Lambert function), representing the nucleation of unfolded units, and (iii) a rigorous demonstration of the equivalence of  the Gibbs and Helmholtz ensembles for the zipper model (nucleation and coalescence peaks disappear as length increases). These results, based on the stationary phase approximation, properly describe the transformation processes in one-dimensional objects (macromolecules or solid nanowires) in terms of thermal fluctuations and mechanical actions applied to the system. Some examples of quantitative comparisons are shown for the microstructural evolution in metallic and semiconductor nanowires.

\backmatter

%\bmhead{Supplementary information}
%
%If your article has accompanying supplementary file/s please state so here. 
%
%Authors reporting data from electrophoretic gels and blots should supply the full unprocessed scans for key as part of their Supplementary information. This may be requested by the editorial team/s if it is missing.
%
%Please refer to Journal-level guidance for any specific requirements.

%\bmhead{Acknowledgments}

\section*{Declarations}
%
%Some journals require declarations to be submitted in a standardised format. Please check the Instructions for Authors of the journal to which you are submitting to see if you need to complete this section. If yes, your manuscript must contain the following sections under the heading `Declarations':
%
\begin{itemize}
\item LB, GF and GP have been supported by the Italian Ministry MIUR-PRIN project Mathematics of active materials: From mechanobiology to smart devices (2017KL4EF3) and by `Gruppo Nazionale per la Fisica Matematica' (GNFM) under `Istituto Nazionale di Alta Matematica' (INdAM). GP and GF are supported by the Italian Ministry MISE through the project RAEE SUD-PVP. GF is also supported by `Istituto Nazionale di Fisica Nucleare' (INFN) through the project QUANTUM, by the FFABR research grant (MIUR) and the PON `S.I.ADD'. AC and SG have been supported by Central Lille and Region Hauts-de-France under project MiBaMs.

\item All authors certify that they have no affiliations with or involvement in any organization or entity with any financial interest or non-financial interest in the subject matter or materials discussed in this manuscript.

%\item Ethics approval 
%\item Consent for publication
\item  All data generated or analysed during this study are included in this published article

\item All authors contributed to the study conception and design. Numerical procedures were performed by AC and LB. Theoretical developments were coordinated by GF, GP and SG. All authors read and approved the final manuscript.
\end{itemize}
%
%\noindent
%If any of the sections are not relevant to your manuscript, please include the heading and write `Not applicable' for that section. 

%%%===================================================%%
%%% For presentation purpose, we have included        %%
%%% \bigskip command. please ignore this.             %%
%%%===================================================%%
%\bigskip
%\bigskip\noindent
%Springer journals and proceedings: \url{https://www.springer.com/gp/editorial-policies}
%
%\bigskip\noindent
%Nature Portfolio journals: \url{https://www.nature.com/nature-research/editorial-policies}
%
%\bigskip\noindent
%\textit{Scientific Reports}: \url{https://www.nature.com/srep/journal-policies/editorial-policies}
%
%\bigskip\noindent
%BMC journals: \url{https://www.biomedcentral.com/getpublished/editorial-policies}
%\end{flushleft}

\begin{appendices}

\section{Non-local behavior: relation between the next to nearest neighbor (NNN) interaction strategy and the Ising scheme}
\label{appA}

\begin{figure}[t]
    \centering
    \includegraphics[scale=0.9]{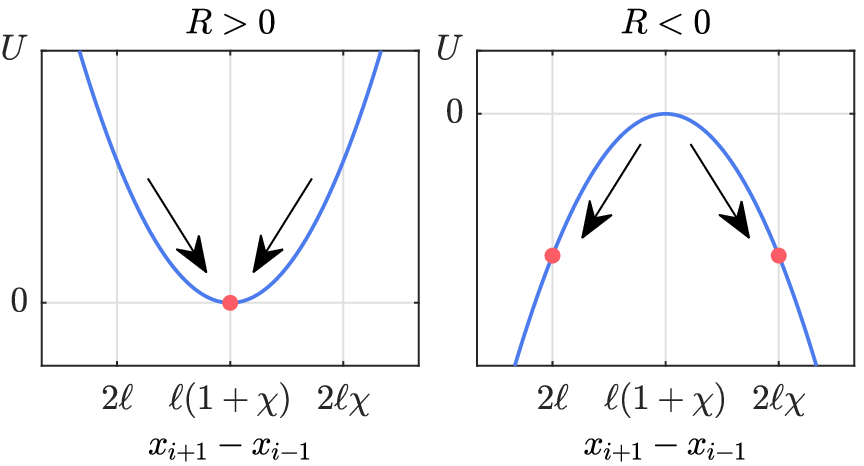}
    \caption{Potential energy $U$ of a linear spring representing an arbitrary NNN element. The two cases correspond to the antiferromagnetic ($R>0$) and ferromagnetic ($R<0$) behaviors of the chain with the NNN linear elastic elements.}
    \label{fig:FAFB}
\end{figure}

The non-local interactions in discrete elastic chains are typically introduced through next to nearest neighbor (NNN) elastic elements \cite{p1,aes}. Under specific hypotheses, we prove here that the scheme with the NNN elements can be reconducted to a typical Ising model, widely adopted in previous studies on similar topics~\cite{maka,PREnew}. To begin with, we consider a chain of bistable units with additional NNN linear springs to model non-local effects~\cite{aes}. The Hamiltonian reads 
\begin{align}
\nonumber
H =& \sum_{i=1}^{N}\left\lbrace Q(S_i)+\frac{K(S_i)\ell^2}{2}\left[\lambda_i-\lambda_0(S_{i})\right]^2\right\rbrace\\
&+\sum_{i=1}^{N-1}\frac{R\ell^2}{2}[\lambda_{i+1}+\lambda_{i}-\lambda_0(-1)-\lambda_0(+1)]^2,
\end{align}
where $R$ is the elastic constant of the NNN springs, and the other quantities are defined in Section~\ref{Isingmodel}. In particular, we name here $H_0$ the energy corresponding to the bistable nearest neighbor (NN) elements and $H_I$ the energy of the NNN linear springs, such that $H=H_0+H_I$. We assume that the equilibrium length of the NNN elements is fixed at $L(-1)+L(+1) = (1+\chi)\ell$, and we introduce two different behaviors: an antiferromagnetic one when $R>0$, and a ferromagnetic one when $R<0$. In the former case ($R>0$), two adjacent units entail a lower energy if they are in two different states (folded and unfolded) while, in the latter one ($R<0$),  two adjacent units result in a lower energy if they are in the same state (either both folded or both unfolded), as shown in Fig.\ref{fig:FAFB}.

%--------------------------------------------------------
%
%
%--------------------------------------------------------

Supposing that the elastic constants $k$ and $\alpha k$ are sufficiently large, the Hamiltonian $H_I$ can be simplified by assuming that the lengths of the units can be approximated with the equilibrium lengths of the explored wells. Then, by using the relations $\lambda_{i+1}\simeq\lambda_0(S_{i+1})$ and $\lambda_{i}\simeq\lambda_0(S_{i})$, we get
\begin{align}
\lambda_{i+1}+\lambda_{i} \simeq\lambda_0(S_{i+1})+\lambda_0(S_{i})
=-\frac{S_{i+1}}{2}(1-\chi)-\frac{S_{i}}{2}(1-\chi)+(1+\chi).
\end{align}
Thus, the interaction Hamiltonian becomes
\begin{align}
\nonumber
H_I&\simeq\sum_{i=1}^{N-1}\frac{R\ell^2}{2}\left[-\frac{(1-\chi)}{2}(S_{i+1}+S_i)\right]^2\\
&= \frac{R}{4}\ell^2(1-\chi)^2(N-1) - J\sum_{i=1}^{N-1}S_{i+1}S_i,
\end{align}
where
\begin{align}
 J=-\frac{R}{4}\ell^2(1-\chi)^2.
\end{align}
This proves that we can approximate the behavior of the chain with the NNN elements by means of a classical Ising chain. We observe that when $J>0$ ($R<0$) we are in a ferromagnetic case, and when $J<0$ ($R>0$) in the antiferromagnetic one. For this reason, in this work we adopted the overall Hamiltonian given by Eq.\eqref{overallH}, where we neglected the irrelevant constant in $H_I$.

\section{Helmholtz and Gibbs ensembles in the Ising Model}
\label{appB}

In this appendix we show the details of the calculation of the Gibbs and Helmholtz partition functions.

%-------------------------------------------
\subsection{Gibbs ensemble}

For the Gibbs ensemble, using Eq.\eqref{overallHa}, the partition function can be evaluated as
\begin{equation}
\begin{split}
	Z_G(\tilde{f})	=\,\,&\ell^N\displaystyle{\sum_{\{S_i\}}}\int_{\mathbb{R}^{N}} e^{-\tilde{\beta}(\tilde{H}-\tilde{f}\sum_{i=1}^N\lambda_i)}\mathrm{d}\lambda_1 \dots\mathrm{d}\lambda_N\\
	=\,\,&\ell^N\displaystyle{\sum_{\{S_i\}}} e^{-\tilde{\beta}\left(\sum_{i=1}^{N} \tilde{Q}(S_i)-\sum_{i=1}^{N-1} S_iS_{i+1}\right)} \\
	&\times\displaystyle{\int_{\mathbb{R}^{N}}}e^{ -\tilde{\beta}\sum_{i=1}^{N}\left[\frac{\tilde{K}(S_i)}{2}\left(\lambda_i -\lambda_0(S_{i})\right)^2- \tilde{f}\lambda_i\right]} \mathrm{d}\lambda_1\dots\mathrm{d}\lambda_N,
\label{gpfapp}
\end{split}
\end{equation}
where $\tilde{f}$ is the dimensionless force applied to the last unit of the chain.
Each sum on $S_i$ ($i=1,\dots,N$) must be interpreted as a sum over the values $+1$ and $-1$.  By a Gaussian integration we obtain
\begin{equation}
Z_G(\tilde{f})	=\ell^N\sum_{\{S_i\}}e^{\tilde{\beta}\sum_{i=1}^{N-1}S_iS_{i+1}}\prod_{i=1}^{N}\sqrt{\frac{2\pi}{\tilde{\beta} \tilde{K}(S_i)}}e^{\tilde{\beta}\left(\frac{\tilde{f}^2}{2\tilde{K}(S_i)}+\lambda_0(S_{i})\tilde{f}-\tilde{Q}(S_i)\right)}. 
\end{equation}
We can define
\begin{equation}
c(S_i)=\sqrt{\frac{2\pi}{\tilde{\beta} \tilde{K}(S_i)}}e^{\tilde{\beta}\left(\frac{\tilde{f}^2}{2\tilde{K}(S_i)}+\lambda_0(S_{i})\tilde{f}-\tilde{Q}(S_i)\right)}, 
\label{cs}
\end{equation}
so that we obtain 
\begin{equation}
\begin{split}
Z_G(\tilde{f})=	\ell^N\displaystyle{\sum_{\{S_i\}}}  \sqrt{c(S_1)}\left[\prod_{i=1}^{N-1}e^{\tilde{\beta} S_i S_{i+1}} 
		\sqrt{c(S_i)c(S_{i+1})}\right]\sqrt{c(S_N)}.
\end{split}
\end{equation}
It order to explicitly evaluate the summation we can use the transfer matrix method \cite{baxter}. We obtain
\begin{equation}\label{gibbsT}
Z_G(\tilde{f})=\boldsymbol{w}^\intercal\boldsymbol{T}^{N-1}\boldsymbol{w},
\end{equation}
where we have defined the transfer matrix $\boldsymbol{T}$ and the vector $\boldsymbol{w}$ (taking care of the boundary conditions) as follows
\begin{equation}\label{TM}
\begin{split}
			\boldsymbol{T}&=\begin{bmatrix}
			e^{\tilde{\beta} }c_-						& e^{-\tilde{\beta} }\sqrt{c_+c_-} \vspace{0.2cm}\\
			e^{-\tilde{\beta} }\sqrt{c_+c_-} 	& e^{\tilde{\beta} }c_+
					\end{bmatrix}, \\
					 &\qquad\qquad\boldsymbol{w}=\begin{pmatrix}\sqrt{c_-}\vspace{0.2cm}\\
			\sqrt{c_+ }
			 \end{pmatrix},
\end{split}
\end{equation}
with, see Eq.\eqref{cs},
\begin{equation}
c_+\triangleq c(+1), \quad c_-\triangleq c(-1).
\label{cpm}
\end{equation}
By using the standard matrix functions theory \cite{gant,lanc}, we have
\begin{equation}\label{Texp}
\boldsymbol{T}^{N-1}=\frac{{\hat \lambda_1}^{N-1}-{\hat \lambda_2}^{N-1}}{{\hat \lambda_1}-{\hat \lambda_2}}\boldsymbol{T}+\frac{{\hat \lambda_1}{\hat \lambda_2}^{N-1}-{\hat \lambda_1}^{N-1}{\hat \lambda_2}}{{\hat \lambda_1}-{\hat \lambda_2}}\boldsymbol{I},
\end{equation}
where $\boldsymbol{I}$ is the $2\times 2$ identity matrix and $\hat \lambda_{1,2}$ are the eigenvalues of $\boldsymbol{T}$, namely
\begin{eqnarray}\label{eigvalT}
\hat \lambda_{1,2} &=&\frac{e^{\tilde{\beta}}}{2}  \left[c_+ + c_- \pm  \sqrt{ \left(c_+-c_-\right)^2+4 c_+c_-e^{-4\tilde{\beta}}} \right]\nonumber\\
&=&\frac{e^{\tilde{\beta}}}{2}  \left(c_+ + c_- \pm  \sqrt{ \Delta} \right).
\end{eqnarray}
In Eq.(\ref{eigvalT}), ${\hat \lambda_1}$ (${\hat \lambda_2}$) corresponds to the $+$ ($-$) sign and we have also defined 
\begin{equation}\label{delta}
\Delta=\left(c_+-c_-\right)^2+4 c_+c_-e^{-4\tilde{\beta}}.
\end{equation}
By substituting $\hat \lambda_{1,2}$ into Eq.\eqref{gibbsT} and Eq.\eqref{Texp}, we get the partition function given in Eq. \eqref{ZGwithlambdas}.
%

%-------------------------------------------
\subsection{Helmholtz  ensemble}

We consider here the case with fixed $x_N$ (isometric condition), described by the Helmholtz ensemble, and evaluate the canonical partition function $Z_H(x_N)$.
The partition functions in the Gibbs and Helmholtz ensembles are linked by a Laplace transform \cite{weiner} 
\begin{eqnarray}
Z_G(f)=\int_{-\infty}^{+\infty}Z_H(x_N)\,e^{\beta f x_N}\mathrm{d}x_N.
\label{laplace}
\end{eqnarray}
Thus, one can write $Z_H(x_N)$, using the change of variable $f\to - i \omega/\beta $, as an inverse Fourier transform in the complex plane
\begin{eqnarray}
\label{Z_H_Lagrange}
Z_H(x_N)=\frac{1}{2\pi}\int_{-\infty}^{+\infty}Z_G\left(-\frac{i\omega}{\beta}\right)\,e^{ i\omega x_N } \mathrm{d}\omega.
\end{eqnarray}
To simplify the notation and perform the calculation we define
\begin{equation}
\delta_{1,2}=e^{-\tilde{\beta}}\,\hat \lambda_{1,2}=\frac{1}{2}  \left(c_+ + c_- \pm  \sqrt{ \Delta} \right).
\label{deltas}
\end{equation}
Thus, the Gibbs partition function can be written as
\begin{equation}
Z_G(\tilde{f})	=\frac{(\ell e^{\tilde{\beta}})^N}{2\cosh\tilde{\beta}} \left[\delta_1^N \left(1+e^{-2\tilde{\beta}}\frac{\delta_1+\delta_2}{\delta_1-\delta_2} \right)+\delta_2^N\left(1-e^{-2\tilde{\beta}}\frac{\delta_1+\delta_2}{\delta_1-\delta_2}\right)\right].
\label{ZGwithdeltas}
\end{equation}
By using the Newton binomial rule to expand the powers $\delta_1^N$ and $\delta_2^N$, we find
\begin{equation}
\begin{split}
Z_G(\tilde{f})	=&\left(\frac{\ell e^{\tilde{\beta}}}{2}\right)^N\frac{1}{2\cosh\tilde{\beta}} \left\{\sum_{k=0}^N \binom{N}{k}(c_++c_-)^{N-k}\Delta^{\frac{k}{2}} \left[1+(-1)^k \right]\right.\\
		&\left.+\sum_{k=0}^N\binom{N}{k}(c_++c_-)^{N-k+1}e^{-2\tilde{\beta}}\Delta^{\frac{k-1}{2}} \left[1-(-1)^k\right]\right\rbrace.
\end{split}
\label{ZGwithdeltas2}
\end{equation}
Here, we can separate the even and odds terms as follows 
\begin{equation}
\sum_{k=0}^Na_k=\sum_{k=0}^{\left[ \frac{N}{2}\right] }a_{2k}+\sum_{k=0}^{\left[ \frac{N-1}{2}\right] }a_{2k+1},
\end{equation}
where the square brackets in the sums stand for the floor function defined as $\left[ x\right] =\max \left\lbrace n\in\mathbb{Z}\vert n\leq x \right\rbrace $.
Then, we get 
\begin{equation}
\begin{array}{ll}
Z_{G}(\tilde{f})= 	& \frac{(\ell e^{\tilde{\beta}})^N}{2^{N}\cosh \tilde{\beta}}\left\{\sum_{k=0}^{\left[\frac{N}{2}\right]}\binom{N}{2k}(c_++c_-)^{N-2k}\Delta^k\right. \\
		&+\left. e^{-2\tilde{\beta}}\sum_{k=0}^{\left[ \frac{N-1}{2}\right]}\binom{N}{2k+1}(c_++c_-)^{N-2k}\Delta^{k}\right\}.
\end{array}
\end{equation}
We can further develop the powers $\Delta^k$ and $(c_++c_-)^{N-2j}$ through the Newton binomial rule, obtaining
\begin{equation}
\begin{array}{ll}
Z_G(\tilde{f})	=&\frac{(\ell e^{\tilde{\beta}})^N}{2^{N}\cosh \tilde{\beta}}\Biggl\{\sum_{k=0}^{\left[\frac{N}{2}\right]}\sum_{j=0}^k\sum_{s=0}^{N-2j}\binom{N}{2k}\binom{k}{j}\binom{N-2j}{s}\Biggr.\\
	&\times c_-^{N-j-s}c_+^{j+s}(-1)^j4^{j}\left(1-e^{-4\tilde{\beta}}\right)^j \\
		&+e^{-2\tilde{\beta}}\sum_{k=0}^{\left[\frac{N-1}{2}\right]}\sum_{j=0}^k\sum_{s=0}^{N-2j}\binom{N}{2k+1}\binom{k}{j}\binom{N-2j}{s}\\
	&\Biggl.\times c_-^{N-j-s}c_+^{j+s}(-1)^j4^{j}\left(1-e^{-4\tilde{\beta}}\right)^j\Biggr\}. \end{array}
\end{equation}
Finally, Eq. \eqref{Z_H_Lagrange} can be evaluated, yielding the canonical partition function within the Helmholtz ensemble given in Eqs.(\ref{zhie}) and (\ref{wjs}).

\end{appendices}




\end{document}